\documentclass{article}
\usepackage{graphicx} % Required for inserting images
\usepackage{a4}
\usepackage{amsmath}
\usepackage{amssymb}
\usepackage[
  colorlinks=true
  ,urlcolor=blue
  ,anchorcolor=blue
  ,citecolor=blue
  ,filecolor=blue
  ,linkcolor=blue
  ,menucolor=blue
  ,linktocpage=true
  ,pdfproducer=medialab
  ,pdfa=true
]{hyperref}
\usepackage[capitalize]{cleveref}

\begin{document}
% Header ---------------------------------------------------
\hspace*{\fill} DESY-24-088

\begin{center}{\Large \bf 
A Benchmarking of QCD evolution at Approximate N$^3$LO
%aN$^3$LO Splitting Function Benchmarking
}\end{center}

\vspace{0.3cm}

\noindent A.~Cooper-Sarkar$^a$, T.~Cridge$^b$, F. Giuli$^c$, L.~A.~Harland-Lang$^d$, 
F.~Hekhorn$^{e,f}$, J.~Huston$^g$, G.~Magni$^{h,i}$, S.~Moch$^j$, R.S.~Thorne$^d$\\

\vspace{0.3cm}

\noindent $^a$ Department of Physics, University of Oxford, Oxford, OX1 3RH, UK\\
$^b$ Deutsches Elektronen-Synchrotron DESY, Notkestr. 85, 22607 Hamburg, Germany\\
$^c$ CERN, CH-1211 Geneva, Switzerland\\
$^d$ Department of Physics and Astronomy, University College London, London, WC1E 6BT, UK \\
$^e$ University of Jyvaskyla, Department of Physics, P.O. Box 35, FI-40014 University of Jyvaskyla, Finland\\
$^f$ Helsinki Institute of Physics, P.O. Box 64, FI-00014 University of Helsinki, Finland\\
$^g$ Department of Physics and Astronomy, Michigan State University, East Lansing, MI 48824, USA\\
$^h$ Department of Physics and Astronomy, Vrije Universiteit, NL-1081 HV Amsterdam\\
$^i$ Nikhef Theory Group, Science Park 105, 1098 XG Amsterdam, The Netherlands\\
$^j$ II. Institute for Theoretical Physics, Hamburg University, D-22761 Hamburg, Germany

\date{March 2024}

\begin{abstract}
    \noindent We present a detailed benchmarking of different treatments of the QCD evolution of unpolarized PDFs at approximate ${\rm N}^3$LO (a${\rm N}^3$LO) order in the QCD coupling. Namely, the implementations in the public a${\rm N}^3$LO releases of the MSHT and NNPDF global PDF fitters, as well as that of the theoretical FHMRUVV collaboration are compared. This follows the same procedure  as in previous benchmarking exercises at lower orders, that is by considering the impact of this evolution on a set of simple toy PDFs. Excellent agreement between the MSHT and NNPDF results is found at NNLO, and at  a${\rm N}^3$LO when the same (FHMRUVV) implementation of the splitting functions is used. In addition, in the data region only small differences between these is seen when their original approximations are used for the splitting functions. The origin of these differences, and the larger ones observed at lower $x$, is well understood in terms of the differences in the approximate splitting functions. Good convergence is also observed between the exact and truncated solution methods for the DGLAP evolution equations. Overall, this provides confidence in the precision of the existing implementation of PDF evolution at  ${\rm N}^3$LO.
\end{abstract}

%\maketitle

\section{Introduction}

The scale dependence of parton distribution functions (PDFs) is governed by QCD evolution equations whose kernels, the splitting functions, are accessible within QCD perturbation theory.
For about 20 years next-to-next-to-leading order (NNLO) has been the standard for precision predictions in perturbative QCD, including studies of the proton structure and fits of PDFs.
For the solution of the integro-differential equations describing parton evolution in QCD, 
different codes have been developed within the community, making it necessary to check that there is consistency between the evolution codes used by the various groups. 
To this end a set of benchmark tables for parton evolution up to NNLO accuracy was produced in 
\cite{Giele:2002hx,Dittmar:2005ed} in order to provide
a means of checking the (numerical) accuracy of any evolution code.

Given the increasing experimental precision and the challenge to push the accuracy of QCD perturbation theory by one quantum loop beyond the state-of-the-art, it is now necessary to extend the benchmarking of QCD evolution codes to the next order, i.e.\ next-to-NNLO (N$^3$LO) in perturbative QCD.
In this document, we provide a new set of benchmark tables for the evolution of unpolarized PDFs.\footnote{A short summary has been published as part of the activities of the “Standard Model” working group for the “Physics at TeV Colliders” workshop (Les Houches, France, 12–30 June, 2023)~\cite{Andersen:2024czj}.
}
As the required QCD splitting functions are not yet fully known at this order, we resort to approximations which are valid within a restricted kinematic range. The outcomes of this study are
\begin{itemize}
    \item benchmarking of available codes for parton evolution to N$^3$LO.
    \item discussion and comparisons of different approximations for the QCD splitting functions at N$^3$LO.
\end{itemize}

We provide a brief summary of the theoretical status of unpolarized parton evolution in QCD in Sec.~\ref{sec:theory}, including a list of available tools and codes. 
We also discuss the different approximations for the N$^3$LO splitting functions.
The results and tables with benchmark numbers are listed in Sec.~\ref{sec:results}.
We finish with conclusions and an outlook in Sec.~\ref{sec:outlook}.

\section{Evolution equations and their solutions}
\label{sec:theory}
\subsection{The renormalization group equations}

The scale evolution of the parton distributions $f_p (x,\mu_{\rm f}^2) \equiv p(x,\mu_{\rm f}^2)$,
where $p = q_i, \,\bar{q}_i\, , g$ with $i = 1, \,\ldots , n_f $,
is governed by the $2n_f\! +\! 1$ coupled integro-differential equations 
\begin{equation}
\label{eq:evol}
 \frac {d\, f_p(x,\mu_{\rm f}^2)} {d \ln \mu_{\rm f}^2} \: = \: 
 \sum_{l=0}^m \, a_{\rm s}^{l+1}(\mu_{\rm r}^2)\:
 {\cal P}_{pp'}^{(l)}\bigg(x,\frac{\mu_{\rm f}^2}{\mu_{\rm r}^2}\bigg) \otimes f_{p'}(x,\mu_{\rm f}^2) \, ,
\end{equation}
where summation over $p'$ is understood. 
The factorization and renormalization scales are denoted $\mu_{\rm f}$ and $\mu_{\rm r}$, and
$\otimes$ is the standard Mellin convolution in the fractional-momentum variable $x$.
The scale dependence of the strong coupling reads 
\begin{equation}
\label{eq:beta}
  \frac{d\, a_{\rm s}}{d \ln \mu_r^2} \: = \: \beta^{}_{\rm N^mLO}(a_{\rm s}) \,=\, - \sum_{l=0}^m \, a_{\rm s}^{l+2} \,\beta_l \, ,
\end{equation}
where we abbreviate $a_{\rm s} \equiv \alpha_{\rm s}/(4\pi)$.
The $\beta$-function coefficients $\beta_l$ are known to five loops in QCD~\cite{Baikov:2016tgj,Herzog:2017ohr,Chetyrkin:2017bjc,Luthe:2017ttg}, which allows for the solution 
of \cref{eq:beta} through N$^4$LO for the $a_{\rm s}$ evolution.
%csm The benchmark computations require the N$^3$LO running of $a_{\rm s}$, so that \cref{eq:beta} is truncated accordingly.
  
The general splitting functions ${\cal P}^{(l)}$ in \cref{eq:evol}
reduce to the simpler expressions $P^{(l)}(x)$ at equal scales $\mu_{\rm r} = \mu_{\rm f}$.
Up to N$^3$LO the corresponding relations read
\begin{eqnarray}
\label{eq:split-mur-muf}
  {\cal P}^{(0)} \bigg(x, \frac{\mu_{\rm f}^2}{\mu_{\rm r}^2} \bigg) 
  &\! =\! & P^{(0)}(x) \, , \nonumber \\
  {\cal P}^{(1)} \bigg(x, \frac{\mu_{\rm f}^2}{\mu_{\rm r}^2} \bigg)
  &\! =\! & P^{(1)}(x) 
  - \beta_0 P^{(0)}(x)\, \ln \frac{\mu_{\rm f}^2}{\mu_{\rm r}^2} \, , \nonumber\\
  {\cal P}^{(2)} \bigg(x, \frac{\mu_{\rm f}^2}{\mu_{\rm r}^2} \bigg)
  &\! =\! & P^{(2)}(x) - \bigg\{ \beta_1 P^{(0)}(x) + 2\beta_0 P^{(1)}(x) 
    \bigg\} \ln \frac{\mu_{\rm f}^2}{\mu_{\rm r}^2} 
    + \beta_0^2 P^{(0)}(x)\, \ln^2 \frac{\mu_{\rm f}^2}{\mu_{\rm r}^2} \, , \\
  {\cal P}^{(3)} \bigg(x, \frac{\mu_{\rm f}^2}{\mu_{\rm r}^2} \bigg)
  &\! =\! & P^{(3)}(x) 
    - \bigg\{ \beta_2 P^{(0)}(x)  + 2 \beta_1 P^{(1)}(x) + 3 \beta_0 P^{(2)}(x) \bigg\} \ln \frac{\mu_{\rm f}^2}{\mu_{\rm r}^2}  \nonumber \\
    &&+  \bigg\{  \frac{5}{2} \beta_0 \beta_1 P^{(0)}(x) + 3 \beta_0^2 P^{(1)}(x)\bigg\} \ln^2 \frac{\mu_{\rm f}^2}{\mu_{\rm r}^2}
    -  \beta_0^3 P^{(0)}(x) \ln^3 \frac{\mu_{\rm f}^2}{\mu_{\rm r}^2} \, .
\nonumber
\end{eqnarray}
The splitting functions are currently known completely up to NNLO, i.e.\ analytic expression for $P_{pp'}^{(2)}(x)$ are available~\cite{Moch:2004pa,Vogt:2004mw}.
At N$^3$LO, this is also the case for non-singlet splitting functions in the planar limit~\cite{Moch:2017uml} as well as for some parts of the functions $P_{pp'}^{(3)}(x)$ 
proportional to powers of $n_f$~\cite{Davies:2016jie,Gehrmann:2023cqm,Gehrmann:2023iah,Falcioni:2023tzp}.
In addition, there are a number of Mellin moments available for all splitting functions at 
N$^3$LO~\cite{Moch:2018wjh,Moch:2021qrk,Falcioni:2023luc,Falcioni:2023vqq,Moch:2023tdj,Falcioni:2024xyt} (at least the first five moments, often ten or even more for specific colour factors).
This information, taken together with knowledge on the functional form of the splitting functions in the limits $x\to 1$ and $x \to 0$, to be discussed below,  serves as the basis for approximations to all N$^3$LO splitting functions, allowing for a solution of \cref{eq:evol} to this order.

\subsection{Heavy-quark treatment}
\label{sec:hq}

The renormalization group equations, \cref{eq:evol,eq:beta}, are viable for a fixed
number of flavors $n_f$ and so we need to specify the prescription for changing
$n_f$. The transition $n_f \to n_f + 1$ is made at the scales of the heavy-quark pole masses,
\begin{equation}
\label{eq:HQ-mpole}
  m_{c} \: = \: \sqrt{2} \mbox{ GeV}^2      \: , \quad 
  m_{b} \: = \: 4.5 \mbox{ GeV}^2  \: . % \quad
 % m_{t} \: = \: 175 \mbox{ GeV}^2  \:\: .
\end{equation}
The matching conditions for the strong coupling in the transition $n_f \to n_f + 1$
derive from the decoupling formulae, given by
\begin{equation}
\label{eq:alphas-matching}
  a_{\rm s}^{\, (n_f+1)}(k_{\rm r} m_h^2) \: = \: 
  a_{\rm s}^{\, (n_f)}(k_{\rm r} m_h^2) + \sum_{n=1}^{m} \, 
  \Big( a_{\rm s}^{\, (n_f)}(k_{\rm r} m_h^2) \Big)^{n+1} 
  \sum_{l=0}^{n} \, c^{}_{n,l} \, \ln k_{\rm r} \:\: .
\end{equation}
The matching coefficients $c^{}_{n,l}$ are known to four loops~\cite{Schroder:2005hy,Chetyrkin:2005ia}, so that  five-loop running can be obtained consistently from \cref{eq:beta}, taking into account threshold effects. The benchmark computations presented here use running $a_{\rm s}^{\, (n_f)}$ at N$^{3}$LO with the four-loop $\beta$-function.

The matching conditions at these thresholds $\mu_{\rm f}^2 = m_{h}^2$, $h = c,\, b,\, t$
for the PDFs involve the operator matrix elements (OMEs) with heavy-quarks, $A_{pp'}(x)$, which are accessible in perturbation theory in an expansion in $a_{\rm s}$ with 
coefficients $A_{pp'}^{(l)}$, known~\cite{Buza:1995ie} to two loops, i.e.\ $A_{pp'}^{(2)}$.
For the benchmark computations, we apply these conditions up to NNLO(=N$^{m=2}$LO) and 
they read~\cite{Buza:1998wv} ($\delta_{m2}=1$ at N$^{m=2}$LO below)
\begin{equation}
\label{eq:HQ-light-quarks}
  l_i^{\,(n_f+1)}(x,m_h^2) \: = \:  l_i^{\,(n_f)}(x,m_h^2) + 
  \delta_{m2} \: a_{\rm s}^2\: A^{NS,(2)}_{qq,h}(x) \otimes 
  l_i^{\, (n_f)}(x,m_h^2) \, ,
\end{equation}
where $l = q,\, \bar{q}$ and $i = 1,\ldots n_f$, and
\begin{eqnarray}
\label{eq:HQ-heavy-quarks}
  g^{\, (n_f+1)}(x,m_h^2) \:\:\: &\! = \!\! & 
    g^{\, (n_f)}(x,m_h^2) + \nonumber \\[0.5mm] & & 
    \delta_{m2} \, a_{\rm s}^2\, \Big[  
    A_{gq,h}^{S,(2)}(x) \otimes \Sigma^{\, (n_f)}(x,m_h^2) +
    A_{gg,h}^{S,(2)}(x) \otimes g^{\, (n_f)}(x,m_h^2) \Big ] \, ,
  \\[0.5mm]
  (h+\bar{h})^{\, (n_f+1)}(x,m_h^2) \! &\! =\!\! & 
    \delta_{m2} \, a_{\rm s}^2\: \Big [ 
    \tilde{A}_{hq}^{S,(2)}(x) \otimes \Sigma^{\, (n_f)}(x,m_h^2) +
    \tilde{A}_{hg}^{S,(2)}(x) \otimes g^{\, (n_f)}(x,m_h^2) \Big  ] \, ,
  \quad \nonumber
\end{eqnarray}
where $h = \bar{h}$ and $\Sigma^{\,(n_f)} \equiv \sum_{i=1}^{n_f} (q_i + \bar{q}_i)$.
%A_qq^h, A_qg^h %A_gq %A_qq^ps
The N$^3$LO QCD corrections to massive OMEs are also available~\cite{Ablinger:2010ty,Ablinger:2014lka,Ablinger:2014nga}. Those for $\tilde{A}_{hg}^{S,(3)}$ have recently been completed~\cite{Ablinger:2023ahe,Ablinger:2024xtt}, superseding previous approximations~\cite{Kawamura:2012cr,Alekhin:2017kpj}.

The massive OMEs also provide the transition matrix elements in the variable-flavor number scheme (VFNS), where the transition $n_f \to n_f + 1$ treats one heavy-quark flavor at the time, 
i.e.\ the single-mass case~\cite{Buza:1996wv}. 
At higher orders, both the $c$ and $b$-quark loops appear in massive OMEs~\cite{Ablinger:2017xml,Ablinger:2018brx,Ablinger:2022wbb}.
Accounting for those corrections requires the transition $n_f \to n_f + 2$, i.e.\ 
the two-mass case~\cite{Ablinger:2017err}.
For the benchmark computations, the VFNS transition employs the single-mass case at NNLO only.

\subsection{Available information and tools}
\label{input+tools}

As it is not possible to solve \cref{eq:evol} in closed form beyond LO,
a numerical algorithm has to be applied for which several public codes exist:
\texttt{APFEL}~\cite{Bertone:2013vaa}, \texttt{apfel++}~\cite{Bertone:2017gds},
\texttt{EKO}~\cite{Candido:2022tld}, \texttt{HOPPET}~\cite{Salam:2008qg},
\texttt{PEGASUS}~\cite{Vogt:2004ns}, and \texttt{QCDNUM}~\cite{Botje:2010ay}.
The codes can be split into two different categories according to how they approach
the solution of \cref{eq:evol}, by either taking a direct approach in momentum
fraction space (\texttt{APFEL}, \texttt{apfel++}, \texttt{HOPPET}, \texttt{QCDNUM}) or solving in the conjugate Mellin-space
(\texttt{EKO}, \texttt{PEGASUS}).

In the following we give a short review on the different approximation strategies
applied by the MSHT group~\cite{McGowan:2022nag,Cridge:2023ryv}, the NNPDF collaboration~\cite{Hekhorn:2023gul,NNPDF:2024nan},  and FHMRUVV~\cite{Moch:2017uml,Falcioni:2023luc,Falcioni:2023vqq,Moch:2023tdj,Falcioni:2024xyt}  - for a detailed discussion we refer to the respective references.

\subsubsection{MSHT}

The first attempt to produce an approximate set of N$^3$LO PDFs, incorporating the known information about the N$^3$LO, i.e.\ ${\cal O}(\alpha_{\rm s}^4)$, splitting functions was made by the MSHT group \cite{McGowan:2022nag}. At the time this study took place less information was available than is currently the case. The approximation for the splitting functions was based on the following procedure
(precise details can be found in \cite{McGowan:2022nag}). 

\begin{itemize}
\item Information on the Mellin moments of the splitting functions   \cite{Moch:2017uml,Moch:2021qrk} was used.
\item The full information on the non-singlet splitting function presented in \cite{Moch:2017uml} was used. This includes information at high $x$ from \cite{Dokshitzer:2005bf} as well as information appropriate in the limit of large flavour and colour number. 
\item The available information on leading $\ln(1/x)$ resummations from 
\cite{Fadin:1975cb,Kuraev:1976ge,Lipatov:1976zz,Kuraev:1977fs,Fadin:1998py,Jaroszewicz:1982gr,Ciafaloni:1998gs,Catani:1994sq} was used. 
\end{itemize}
For each splitting function the terms which are known exactly were input, together with a set of basis functions chosen to be concentrated over a variety of $x$ regions. 
For each splitting function one more function than there are known Mellin moments was chosen (i.e.\ a set of 5 basis functions to accompany the 4 moments known at the time for singlet splitting functions and a set of 9 basis functions to accompany the 8 moments for the nonsinglet splitting function). This leaves an unknown parameter,
$\rho$ which reflects the splitting function uncertainty and was chosen to be the coefficient of the most divergent unknown piece of our set of basis functions at small $x$. The variation in this parameter was then constrained by two requirements: for allowed $\rho$ at high $x$, when the other coefficients of basis parameters are determined by matching the full splitting functions to the known Mellin moments, then  the splitting function corrections at N$^3$LO should not be large compare to those at lower orders; at sufficiently small $x$, for a fixed value of $\rho$,  the splitting function is required to be contained within the range of variation 
predicted from exploring a full range of choice of basis functions. Once  a particular choice of basis functions has been made there is then a preferred value of $\rho$ and a variation which incurs a $\chi^2$ penalty, i.e.\ there is a prior based on all the available knowledge about the function and lower orders.  This prior was then modified in a fit to data and a posterior ``best-fit'' value and uncertainty was obtained for the value of $\rho$ for each splitting function. 
The constraints on the splitting functions were applied in moment space in order to determine the values of $\rho$, but the splitting function was then expressed in $x$ space and evolution performed via numerical solution of the renormalization group equations in \cref{eq:evol}. 

This approach was used in both the first release of the approximate N$^3$LO PDFs \cite{McGowan:2022nag} and subsequent work including also QED evolution \cite{Cridge:2023ryv}, examining various aspects of the data used for the high $x$ gluon in the aN$^3$LO PDFs \cite{Cridge:2023ozx}, and the first determination of the strong coupling at aN$^3$LO in a global PDF fit \cite{Cridge:2024exf}.

\subsubsection{NNPDF}

The default program adopted by the NNPDF collaboration for DGLAP evolution is the recently published code \texttt{EKO}~\cite{Candido:2022tld}. 
It focuses on solving \cref{eq:evol} in terms of evolution kernel 
operators (EKOs) $E$ which are independent from the actual PDF $f_p(x,\mu_{\rm f}^2)$,
depending only on the theory setup (such as perturbative orders or the initial scale $\mu_{\rm f,0}^2$):
\begin{equation}
    f_p(\mu_{\rm f}^2) = E_{pp'}(\mu_{\rm f}^2 \leftarrow \mu_{\rm f,0}^2) \otimes f_{p'}(\mu_{\rm f,0}^2)
\end{equation}
The EKOs are computed in Mellin space and transformed to momentum fraction space via
interpolation techniques. \texttt{EKO} provides several strategies for solving the
respective renormalization group equations which are perturbatively equivalent, but
differ by the number of resummed terms.
As EKOs are independent of the boundary PDF values, they are ideally suited for a PDF
fit and as such \texttt{EKO} is integrated into the \texttt{pineline} framework~\cite{Barontini:2023vmr}.

The \texttt{EKO} strategy for approximating the N$^3$LO splitting function $P^{(3)}$, can be summarized as follows:
\begin{itemize}
    \item The available information from soft gluon resummation in the threshold region $x\to 1$~\cite{Moch:2017uml,Davies:2016jie,Dokshitzer:2005bf, Henn:2019swt,Duhr:2022cob,Almasy:2010wn} is used.
    \item The available information from BFKL resummation in the high-energy region $x\to 0$~\cite{Fadin:1975cb,Kuraev:1976ge,Lipatov:1976zz,Kuraev:1977fs,Fadin:1998py,Jaroszewicz:1982gr,Davies:2022ofz,Ball:1999sh,Bonvini:2018xvt} is used.
    \item The available information in the limit of large number of flavors~\cite{Moch:2017uml,Davies:2016jie,Gehrmann:2023cqm,Falcioni:2023tzp} is used.
    \item The available information on the low Mellin moments~\cite{Moch:2017uml,Moch:2021qrk,Falcioni:2023luc,Falcioni:2023vqq,Moch:2023tdj}
is used.
\end{itemize}
All limits are combined in a unique way and a number of sub-leading terms corresponding to the number of known Mellin moments are added. 
Precise details can be found in Section~2 of \cite{NNPDF:2024nan}.
In particular the specific choice of the form of the sub-leading terms is arbitrary and drawn from a pre-defined list.
This freedom represents the uncertainty associated with the partial knowledge on splitting functions and is referred to as Incomplete Higher Order Uncertainty (IHOU).
IHOU are estimated independently from other possible sources of theory errors, such as the incomplete perturbative expansion (referred to as missing higher order uncertainty (MHOU)~\cite{NNPDF:2019ubu,NNPDF:2024dpb}).
%, and they do not depend on the final result of the PDF fit.

\subsubsection{FHMRUVV}

Approximate results for the N$^3$LO splitting functions have been constructed based on 
the available Mellin moments~\cite{Moch:2018wjh,Moch:2021qrk,Falcioni:2023luc,Falcioni:2023vqq,Moch:2023tdj,Falcioni:2024xyt} 
and all known results for the large-$x$ and small-$x$ limits.
The large-$x$ behavior of the ${\overline{\rm MS}}$-scheme diagonal splitting functions reads
\begin{equation}
\label{eq:Pkk-xto1}
  P_{pp,\,x\to 1\,}^{\,(l-1)}(x) \:\: = \:\;
        \frac{A_{p}^{(l)}}{(1-x)_+}
  \,+\, B_{p}^{\,(l)} \, \delta (1\!-\!x)
  \,+\, C_{p}^{\,(l)} \, \ln (1-x)
  \,-\, A_{p}^{(l)} + D_{p}^{\,(l)}
\:\: ,
\end{equation}
where the four-loop cusp anomalous dimensions $A_{p}^{(4)}$~\cite{Henn:2019swt,vonManteuffel:2020vjv}
as well as the coefficents $C_{p}^{\,(4)}$ and $D_{p}^{\,(4)}$~\cite{Moch:2023tdj,Dokshitzer:2005bf} are known analytically, while the 
virtual anomalous dimensions $B_{p}^{\,(4)}$ multiplying $\delta(1-x)$ 
have been determined in approximate form~\cite{Das:2019btv,Das:2020adl}.
For all N$^3$LO splitting functions further large-$x$ constraints are known~\cite{Almasy:2010wn,Soar:2009yh,Vogt:2010cv}  and are also used.
Information on their small-$x$ behavior is also available. 
In case of $P_{gg}^{(3)}$ it is governed by the BFKL logarithms, of which the next-to-leading logarithmic (NLL) correction has been 
calculated \cite{Fadin:1998py,Ciafaloni:1998gs} and transformed to the
${\overline{\rm MS}}$-scheme some time ago~\cite{Ciafaloni:2005cg,Ciafaloni:2006yk} (for $P_{gq}^{(3)}$, see~\cite{Catani:1994sq}). 
Sub-dominant small-$x$ terms for all splitting functions have also been derived~\cite{Davies:2022ofz}.

The procedure is then based on a total of 80 approximations featuring slightly different functional forms 
for the splitting functions $P_{pp'}^{\,(3)}(x)$, consistent with their respective known endpoint behaviour.
These approximations define a range and for the two boundaries of this range 
selected representatives $P_{pp'\,A}^{\,(3)}(x)$ and $P_{pp',\,B}^{\,(3)}(x)$ 
are presented to provide the error bands for $n_f = 3,\,4,\,5$ light flavours.
The quality of this procedure is tested through approximations of the known NNLO splitting functions, demonstrating good convergence and providing information on the range of $x$ for which residual uncertainties due to missing higher Mellin moments are small.

The approximate N$^3$LO splitting functions are valid at large-$x$, with the uncertainties increasing towards smaller values of $x$.
The FHMRUVV results~\cite{Moch:2017uml,Falcioni:2023luc,Falcioni:2023vqq,Moch:2023tdj}
for $P_{qq}^{\,(3)}(x)$ (non-singlet and pure-singlet) and $P_{qg}^{\,(3)}(x)$,
based on ten Mellin moments $N\leq 20$, are reliable approximations down to $x \simeq 10^{-4}$.
The quantities $P_{gq}^{\,(3)}(x)$ and $P_{gg}^{\,(3)}(x)$, based on five Mellin moments $N\leq 10$, are reasonable approximations down to $x \simeq 10^{-3}$.
Their convolution with PDFs will dampen the uncertainty at small-$x$, 
so that one gains one additional order of magnitude on the right hand side of the evolution in \cref{eq:evol} due to fast falling PDFs as $x \to 1$. 
The N$^3$LO approximations to splitting functions convoluted with PDFs are, therefore,  reliable in the range $10^{-5} \lesssim x \leq 1$ ($P_{qq}^{\,(3)}(x)$, $P_{qg}^{\,(3)}(x)$) and $10^{-4} \lesssim x \leq 1$ ($P_{gq}^{\,(3)}(x)$, $P_{gg}^{\,(3)}(x)$), respectively. 
The latest results for $P_{gq}^{\,(3)}(x)$~\cite{Falcioni:2024xyt}, based on ten Mellin moments ($N\leq 20$), have not been used in this benchmark comparison.

\section{Benchmark tables} 
\label{sec:results}

The corresponding lower order benchmark results are available at LO and NLO
in \cite{Giele:2002hx} and at NNLO in \cite{Dittmar:2005ed},
where LO and NLO polarized evolution has also been included.
A number of minor typos have been reported in ~\cite{Candido:2022tld,Diehl:2021gvs} and are listed here for completeness:

\begin{itemize}
\item In \cite{Giele:2002hx}, table headers, the combination $L_{+}$ has to be defined as $L_{+}= 2 (\bar{d} + \bar{u})$.
\item In \cite{Giele:2002hx}, header of Tab.~1 $\alpha_s(\mu_{\rm r}^2=10^4 \mbox{ GeV}^2) = 0.117574$, as pointed out in \cite{Dittmar:2005ed}.
\item In \cite{Giele:2002hx}, Tab.~1, $xs_+(x=0.5, \mu_{\rm f}=10^4 \mbox{ GeV}^2 ) = 7.3137 \cdot 10^{-4}$ for $n_f =4$, and $xb_+(x=10^{-7}, \mu_{\rm f}=10^4 \mbox{ GeV}^2 ) = 4.6071 \cdot 10^{+1}$ for $n_f =3 \dots 5$, differ in the last digit.
\item In \cite{Giele:2002hx}, Tab.~4, the value $xu_v(x=0.7, \mu_{\rm f}=10^4 \mbox{ GeV}^2 ) = 2.0102 \cdot 10^{-2}$,  differs in the last digit.
\item In \cite{Giele:2002hx}, Tab.~3, the values 
$xL_-(x=10^{-5}, \mu_{\rm f}=10^4 \mbox{ GeV}^2 ) = 1.0121 \cdot 10^{-4}$ and 
$xL_-(x=10^{-1}, \mu_{\rm f}=10^4 \mbox{ GeV}^2 ) = 9.8435 \cdot 10^{-3}$, contain wrong exponents.
\item In \cite{Dittmar:2005ed}, Tab.~15,  the values
$xd_v(x=10^{-7}, \mu_{\rm f}=10^4 \mbox{ GeV}^2 ) = 1.0699 \cdot 10^{-4}$ and 
$xg(x=10^{-7}, \mu_{\rm f}=10^4 \mbox{ GeV}^2 ) = 9.9694 \cdot 10^{2}$, contain wrong exponents. 
% \item In \cite{Dittmar:2005ed}, Tab.~16, for polarized evolution, the distribution $L_-$
%  is not vanishing despite the fact that the toy PDF, defined in Eq.~4.57, has $\bar{u} = \bar{d}$. The whole column should be set to 0, since LO evolution does not generate an asymmetry between and $\bar{u}$ and $\bar{d}$.
\end{itemize}

\subsection{Initial conditions}
\label{sec:toy_pdf}

The initial conditions for the reference results are taken from~\cite{Giele:2002hx,Dittmar:2005ed}
with the evolution starting at 
\begin{equation}
\label{eq:muf-start}
  \mu_{\rm f,0}^2 \: = \: 2 \mbox{ GeV}^2 \, ,
\end{equation}
and the parametrizations of input distributions are chosen as
\begin{align}
\label{eq:input}
  xu_v(x,\mu_{\rm f,0}^2)       &= 5.107200\: x^{0.8}\: (1-x)^3  \, ,
    \nonumber \\
  xd_v(x,\mu_{\rm f,0}^2)       &= 3.064320\: x^{0.8}\: (1-x)^4  \, ,
    \nonumber \\
  xg\,(x,\mu_{\rm f,0}^2)       &= 1.700000\, x^{-0.1} (1-x)^5  \, ,
    \\
  x\bar{d}\,(x,\mu_{\rm f,0}^2) &= 0.1939875\, x^{-0.1} (1-x)^6  \, ,
    \nonumber\\
  x\bar{u}\,(x,\mu_{\rm f,0}^2) &= (1-x)\: x\bar{d}\,(x,\mu_{\rm f,0}^2) \, ,
    \nonumber\\
  xs\,(x,\mu_{\rm f,0}^2)       &= x\bar{s}\,(x,\mu_{\rm f,0}^2) 
    \: = \: 0.2\, x(\bar{u}+\bar{d}\,)(x,\mu_{\rm f,0}^2) \, ,
    \nonumber  
\end{align}
where the valence distributions are defined as $q_{i,v} \equiv q_i - \bar{q}_i$
and the value for the running coupling at the input scale is fixed as 
\begin{equation}
\label{eq:alphas-start}
  \alpha_{\rm s}(\mu_{\rm r}^2\! =\! 2\mbox{ GeV}^2) \: = \: 0.35 \:\: .
\end{equation}
These initial conditions are used regardless of the perturbative order of the evolution
and of the ratio of the renormalization and factorization scales, which take
the values
\begin{equation}
\label{eq:muf-mur-scale-var}
  \mu_{\rm r}^2 \: = \: k_{\rm r} \mu_{\rm f}^2 \:\: , \quad\quad 
  k_{\rm r} \: = \:  0.5\, , \:\: 1\, , \:\: 2 \, ,
\end{equation}
except for LO, where this ratio is fixed to unity.

We assume two different setups for the heavy quark treatment discussed in \cref{sec:hq}:
\begin{enumerate}
    \item We assume a fixed number $n_f=4$ of quarks participating in DGLAP evolution and we refer to this as the Fixed Flavor Number scheme (FFNS). We assume that all distributions in \cref{eq:input} are given directly in this scheme and that the charm distribution vanishes at the initial scale $\mu_{\rm f,0}^2$. Similarly, we assume that the boundary condition of the strong coupling \cref{eq:alphas-start} is given in this scheme.
    \item We assume a dynamic range $n_f = 3 \ldots 5$ of participating quark flavors for our Variable Flavor Number scheme (VFNS), where each quark (here charm and bottom) gets activated when the evolution scale matches the quark mass. We assume \cref{eq:input,eq:alphas-start} apply for $n_f=3$, such that an immediate matching procedure is triggered at the beginning of the evolution.
\end{enumerate}

\subsection{Results}

The benchmark values are reported in the same basis adopted in \cite{Dittmar:2005ed}, and we define:
\begin{equation}
    q_v = q - \bar q\, , \quad
    L_{-} = (\bar{d} - \bar{u})\, , \quad
    L_{+} = 2(\bar{d} + \bar{u})\, , \quad
    q_+ = q + \bar{q}  \, .
\end{equation}
In order to simplify the comparison, only NNLO matching conditions are adopted,
although the N$^3$LO coefficients are now available in literature and 
implemented in the evolution tools.
The evolved PDFs are then evaluated at a final scale $\mu_{\rm f}^2=10^4 \mbox{ GeV}^2$, 
which is chosen as representative for LHC processes.
As a preliminary condition to the N$^3$LO benchmark, we have tested that the different evolution codes are 
able to reproduce the NNLO benchmark tables of \cite{Dittmar:2005ed} with an accuracy of $0.01\%$ or below.
We adopt the same notation as in the previous study and write $x^{a} = x \times 10^{a}$.

   \begin{table}[htp]
    \caption{
        Results for the FFNS aN$^3$LO evolution
        for the initial conditions and the input parton distributions
        given in Sec.~\ref{sec:toy_pdf},
        with the FHMRUVV splitting functions approximation and the MSHT20aN3LO code.}
    \label{tab:n3lo_ffns_msht_fhmruvv}
    \begin{center}
    \vspace{5mm}
\begin{tabular}{||c||r|r|r|r|r|r|r|r||}
    \hline \hline
    \multicolumn{9}{||c||}{} \\[-3mm]
    \multicolumn{9}{||c||}{FHMRUVV in MSHT20aN3LO $\, n_f = 4$,
        $\,\mu_{\rm f}^2 = 10^4 \mbox{ GeV}^2$} \\
    \multicolumn{9}{||c||}{} \\[-0.3cm]
    \hline \hline
    \multicolumn{9}{||c||}{} \\[-3mm]
    \multicolumn{1}{||c||}{$x$} &
    \multicolumn{1}{c|} {$xu_v$} &
    \multicolumn{1}{c|} {$xd_v$} &
    \multicolumn{1}{c|} {$xL_-$} &
    \multicolumn{1}{c|} {$xL_+$} &
    \multicolumn{1}{c|} {$xs_v$} &
    \multicolumn{1}{c|} {$xs_+$} &
    \multicolumn{1}{c|} {$xc_+$} &
    \multicolumn{1}{c||}{$xg$} \\[0.5mm]

\hline \hline
\multicolumn{9}{||c||}{} \\[-3mm]
\multicolumn{9}{||c||}{$\mu_{\rm r}^2 = \ \mu_{\rm f}^2$} \\
\multicolumn{9}{||c||}{} \\
\hline \hline
\end{tabular}
\end{center}
\end{table}
    \begin{table}[htp]
    \caption{
        Results for the FFNS aN$^3$LO evolution
        for the initial conditions and the input parton distributions
        given in Sec.~\ref{sec:toy_pdf},
        with the FHMRUVV splitting functions approximation and the NNPDF code.
    }
    \label{tab:n3lo_ffns_fhmruvv}
    \begin{center}
    \vspace{5mm}
    \begin{tabular}{||c||r|r|r|r|r|r|r|r||}
    \hline \hline
    \multicolumn{9}{||c||}{} \\[-3mm]
    \multicolumn{9}{||c||}{FHMRUVV in NNPDF $\, n_f = 4$,
        $\,\mu_{\rm f}^2 = 10^4 \mbox{ GeV}^2$} \\
    \multicolumn{9}{||c||}{} \\[-0.3cm]
    \hline \hline
    \multicolumn{9}{||c||}{} \\[-3mm]
    \multicolumn{1}{||c||}{$x$} &
    \multicolumn{1}{c|} {$xu_v$} &
    \multicolumn{1}{c|} {$xd_v$} &
    \multicolumn{1}{c|} {$xL_-$} &
    \multicolumn{1}{c|} {$xL_+$} &
    \multicolumn{1}{c|} {$xs_v$} &
    \multicolumn{1}{c|} {$xs_+$} &
    \multicolumn{1}{c|} {$xc_+$} &
    \multicolumn{1}{c||}{$xg$} \\[0.5mm]
    
\hline \hline
\multicolumn{9}{||c||}{} \\[-3mm]
\multicolumn{9}{||c||}{$\mu_{\rm r}^2 = \ \mu_{\rm f}^2$} \\
\multicolumn{9}{||c||}{} \\
\hline \hline
\end{tabular}
\end{center}
\end{table}

   \begin{table}[htp]
    \caption{
        Results for the FFNS aN$^3$LO evolution
        for the initial conditions and the input parton distributions
        given in Sec.~\ref{sec:toy_pdf},
        with the MSHT20aN3LO prior splitting functions approximation.}
    \label{tab:n3lo_ffns_mshtprior}
    \begin{center}
    \vspace{5mm}
    \begin{tabular}{||c||r|r|r|r|r|r|r|r||}
    \hline \hline
    \multicolumn{9}{||c||}{} \\[-3mm]
    \multicolumn{9}{||c||}{MSHT20aN3LO prior $\, n_f = 4$,
        $\,\mu_{\rm f}^2 = 10^4 \mbox{ GeV}^2$} \\
    \multicolumn{9}{||c||}{} \\[-0.3cm]
    \hline \hline
    \multicolumn{9}{||c||}{} \\[-3mm]
    \multicolumn{1}{||c||}{$x$} &
    \multicolumn{1}{c|} {$xu_v$} &
    \multicolumn{1}{c|} {$xd_v$} &
    \multicolumn{1}{c|} {$xL_-$} &
    \multicolumn{1}{c|} {$xL_+$} &
    \multicolumn{1}{c|} {$xs_v$} &
    \multicolumn{1}{c|} {$xs_+$} &
    \multicolumn{1}{c|} {$xc_+$} &
    \multicolumn{1}{c||}{$xg$} \\[0.5mm]
    
\hline \hline
\multicolumn{9}{||c||}{} \\[-3mm]
\multicolumn{9}{||c||}{$\mu_{\rm r}^2 = \ \mu_{\rm f}^2$} \\
\multicolumn{9}{||c||}{} \\
\hline \hline
\end{tabular}
\end{center}
\end{table}

   \begin{table}[htp]
    \caption{
        Results for the FFNS aN$^3$LO evolution
        for the initial conditions and the input parton distributions
        given in Sec.~\ref{sec:toy_pdf},
        with the MSHT20aN3LO posterior splitting functions approximation.}
    \label{tab:n3lo_ffns_mshtposterior}
    \begin{center}
    \vspace{5mm}
    \begin{tabular}{||c||r|r|r|r|r|r|r|r||}
    \hline \hline
    \multicolumn{9}{||c||}{} \\[-3mm]
    \multicolumn{9}{||c||}{MSHT20aN3LO posterior $\, n_f = 4$,
        $\,\mu_{\rm f}^2 = 10^4 \mbox{ GeV}^2$} \\
    \multicolumn{9}{||c||}{} \\[-0.3cm]
    \hline \hline
    \multicolumn{9}{||c||}{} \\[-3mm]
    \multicolumn{1}{||c||}{$x$} &
    \multicolumn{1}{c|} {$xu_v$} &
    \multicolumn{1}{c|} {$xd_v$} &
    \multicolumn{1}{c|} {$xL_-$} &
    \multicolumn{1}{c|} {$xL_+$} &
    \multicolumn{1}{c|} {$xs_v$} &
    \multicolumn{1}{c|} {$xs_+$} &
    \multicolumn{1}{c|} {$xc_+$} &
    \multicolumn{1}{c||}{$xg$} \\[0.5mm]
    
\hline \hline
\multicolumn{9}{||c||}{} \\[-3mm]
\multicolumn{9}{||c||}{$\mu_{\rm r}^2 = \ \mu_{\rm f}^2$} \\
\multicolumn{9}{||c||}{} \\
\hline \hline
\end{tabular}
\end{center}
\end{table}

    \begin{table}[htp]
    \caption{
        Results for the FFNS aN$^3$LO evolution
        for the initial conditions and the input parton distributions
        given in Sec.~\ref{sec:toy_pdf},
        with the NNPDF splitting functions approximation.
    }
    \label{tab:n3lo_ffns_nnpdf}
    \begin{center}
    \vspace{5mm}
    \begin{tabular}{||c||r|r|r|r|r|r|r|r||}
    \hline \hline
    \multicolumn{9}{||c||}{} \\[-3mm]
    \multicolumn{9}{||c||}{NNPDF $\, n_f = 4$,
        $\,\mu_{\rm f}^2 = 10^4 \mbox{ GeV}^2$} \\
    \multicolumn{9}{||c||}{} \\[-0.3cm]
    \hline \hline
    \multicolumn{9}{||c||}{} \\[-3mm]
    \multicolumn{1}{||c||}{$x$} &
    \multicolumn{1}{c|} {$xu_v$} &
    \multicolumn{1}{c|} {$xd_v$} &
    \multicolumn{1}{c|} {$xL_-$} &
    \multicolumn{1}{c|} {$xL_+$} &
    \multicolumn{1}{c|} {$xs_v$} &
    \multicolumn{1}{c|} {$xs_+$} &
    \multicolumn{1}{c|} {$xc_+$} &
    \multicolumn{1}{c||}{$xg$} \\[0.5mm]
    
\hline \hline
\multicolumn{9}{||c||}{} \\[-3mm]
\multicolumn{9}{||c||}{$\mu_{\rm r}^2 = \ \mu_{\rm f}^2$} \\
\multicolumn{9}{||c||}{} \\
\hline \hline
\end{tabular}
\end{center}
\end{table}

   \begin{table}[htp]
    \caption{
        Results for the VFNS aN$^3$LO evolution
        for the initial conditions and the input parton distributions
        given in Sec.~\ref{sec:toy_pdf},
        with the FHMRUVV splitting functions approximation and the MSHT20aN3LO code.
    }
    \label{tab:n3lo_vfns_fhmruvv_msht}
    \begin{center}
    \vspace{5mm}
    \begin{tabular}{||c||r|r|r|r|r|r|r|r||}
    \hline \hline
    \multicolumn{9}{||c||}{} \\[-3mm]
    \multicolumn{9}{||c||}{FHMRUVV in MSHT20aN3LO, $\, n_f = 3\ldots 5\,$,
        $\,\mu_{\rm f}^2 = 10^4 \mbox{ GeV}^2$} \\
    \multicolumn{9}{||c||}{} \\[-0.3cm]
    \hline \hline
    \multicolumn{9}{||c||}{} \\[-3mm]
    \multicolumn{1}{||c||}{$x$} &
    \multicolumn{1}{c|} {$xu_v$} &
    \multicolumn{1}{c|} {$xd_v$} &
    \multicolumn{1}{c|} {$xL_-$} &
    \multicolumn{1}{c|} {$xL_+$} &
    \multicolumn{1}{c|} {$xs_+$} &
    \multicolumn{1}{c|} {$xc_+$} &
    \multicolumn{1}{c|} {$xb_+$} &
    \multicolumn{1}{c||}{$xg$} \\[0.5mm]
    
\hline \hline
\multicolumn{9}{||c||}{} \\[-3mm]
\multicolumn{9}{||c||}{$\mu_{\rm r}^2 = \ \mu_{\rm f}^2$} \\
\multicolumn{9}{||c||}{} \\
\hline \hline
\end{tabular}
\end{center}
\end{table}

    \begin{table}[htp]
    \caption{
        Results for the VFNS aN$^3$LO evolution
        for the initial conditions and the input parton distributions
        given in Sec.~\ref{sec:toy_pdf},
        with the FHMRUVV splitting functions approximation and the NNPDF code.
    }
    \label{tab:n3lo_vfns_fhmruvv}
    \begin{center}
    \vspace{5mm}
    \begin{tabular}{||c||r|r|r|r|r|r|r|r||}
    \hline \hline
    \multicolumn{9}{||c||}{} \\[-3mm]
    \multicolumn{9}{||c||}{FHMRUVV in NNPDF, $\, n_f = 3\ldots 5\,$,
        $\,\mu_{\rm f}^2 = 10^4 \mbox{ GeV}^2$} \\
    \multicolumn{9}{||c||}{} \\[-0.3cm]
    \hline \hline
    \multicolumn{9}{||c||}{} \\[-3mm]
    \multicolumn{1}{||c||}{$x$} &
    \multicolumn{1}{c|} {$xu_v$} &
    \multicolumn{1}{c|} {$xd_v$} &
    \multicolumn{1}{c|} {$xL_-$} &
    \multicolumn{1}{c|} {$xL_+$} &
    \multicolumn{1}{c|} {$xs_+$} &
    \multicolumn{1}{c|} {$xc_+$} &
    \multicolumn{1}{c|} {$xb_+$} &
    \multicolumn{1}{c||}{$xg$} \\[0.5mm]
    
\hline \hline
\multicolumn{9}{||c||}{} \\[-3mm]
\multicolumn{9}{||c||}{$\mu_{\rm r}^2 = \ \mu_{\rm f}^2$} \\
\multicolumn{9}{||c||}{} \\
\hline \hline
\end{tabular}
\end{center}
\end{table}

   \begin{table}[htp]
    \caption{
        Results for the VFNS aN$^3$LO evolution
        for the initial conditions and the input parton distributions
        given in Sec.~\ref{sec:toy_pdf},
        with the MSHT20aN3LO prior splitting functions approximation.
    }
    \label{tab:n3lo_vfns_msht_prior}
    \begin{center}
    \vspace{5mm}
    \begin{tabular}{||c||r|r|r|r|r|r|r|r||}
    \hline \hline
    \multicolumn{9}{||c||}{} \\[-3mm]
    \multicolumn{9}{||c||}{MSHT20aN3LO prior, $\, n_f = 3\ldots 5\,$,
        $\,\mu_{\rm f}^2 = 10^4 \mbox{ GeV}^2$} \\
    \multicolumn{9}{||c||}{} \\[-0.3cm]
    \hline \hline
    \multicolumn{9}{||c||}{} \\[-3mm]
    \multicolumn{1}{||c||}{$x$} &
    \multicolumn{1}{c|} {$xu_v$} &
    \multicolumn{1}{c|} {$xd_v$} &
    \multicolumn{1}{c|} {$xL_-$} &
    \multicolumn{1}{c|} {$xL_+$} &
    \multicolumn{1}{c|} {$xs_+$} &
    \multicolumn{1}{c|} {$xc_+$} &
    \multicolumn{1}{c|} {$xb_+$} &
    \multicolumn{1}{c||}{$xg$} \\[0.5mm]
    
\hline \hline
\multicolumn{9}{||c||}{} \\[-3mm]
\multicolumn{9}{||c||}{$\mu_{\rm r}^2 = \ \mu_{\rm f}^2$} \\
\multicolumn{9}{||c||}{} \\
\hline \hline
\end{tabular}
\end{center}
\end{table}

   \begin{table}[htp]
    \caption{
        Results for the VFNS aN$^3$LO evolution
        for the initial conditions and the input parton distributions
        given in Sec.~\ref{sec:toy_pdf},
        with the MSHT20aN3LO posterior splitting functions approximation.
    }
    \label{tab:n3lo_vfns_msht_posterior}
    \begin{center}
    \vspace{5mm}
    \begin{tabular}{||c||r|r|r|r|r|r|r|r||}
    \hline \hline
    \multicolumn{9}{||c||}{} \\[-3mm]
    \multicolumn{9}{||c||}{MSHT20aN3LO posterior, $\, n_f = 3\ldots 5\,$,
        $\,\mu_{\rm f}^2 = 10^4 \mbox{ GeV}^2$} \\
    \multicolumn{9}{||c||}{} \\[-0.3cm]
    \hline \hline
    \multicolumn{9}{||c||}{} \\[-3mm]
    \multicolumn{1}{||c||}{$x$} &
    \multicolumn{1}{c|} {$xu_v$} &
    \multicolumn{1}{c|} {$xd_v$} &
    \multicolumn{1}{c|} {$xL_-$} &
    \multicolumn{1}{c|} {$xL_+$} &
    \multicolumn{1}{c|} {$xs_+$} &
    \multicolumn{1}{c|} {$xc_+$} &
    \multicolumn{1}{c|} {$xb_+$} &
    \multicolumn{1}{c||}{$xg$} \\[0.5mm]
    
\hline \hline
\multicolumn{9}{||c||}{} \\[-3mm]
\multicolumn{9}{||c||}{$\mu_{\rm r}^2 = \ \mu_{\rm f}^2$} \\
\multicolumn{9}{||c||}{} \\
\hline \hline
\end{tabular}
\end{center}
\end{table}

    \begin{table}[htp]
    \caption{
        Results for the VFNS aN$^3$LO evolution
        for the initial conditions and the input parton distributions
        given in Sec.~\ref{sec:toy_pdf},
        with the NNPDF splitting functions approximation.
    }
    \label{tab:n3lo_vfns_nnpdf}
    \begin{center}
    \vspace{5mm}
    \begin{tabular}{||c||r|r|r|r|r|r|r|r||}
    \hline \hline
    \multicolumn{9}{||c||}{} \\[-3mm]
    \multicolumn{9}{||c||}{NNPDF, $\, n_f = 3\ldots 5\,$,
        $\,\mu_{\rm f}^2 = 10^4 \mbox{ GeV}^2$} \\
    \multicolumn{9}{||c||}{} \\[-0.3cm]
    \hline \hline
    \multicolumn{9}{||c||}{} \\[-3mm]
    \multicolumn{1}{||c||}{$x$} &
    \multicolumn{1}{c|} {$xu_v$} &
    \multicolumn{1}{c|} {$xd_v$} &
    \multicolumn{1}{c|} {$xL_-$} &
    \multicolumn{1}{c|} {$xL_+$} &
    \multicolumn{1}{c|} {$xs_+$} &
    \multicolumn{1}{c|} {$xc_+$} &
    \multicolumn{1}{c|} {$xb_+$} &
    \multicolumn{1}{c||}{$xg$} \\[0.5mm]
    
\hline \hline
\multicolumn{9}{||c||}{} \\[-3mm]
\multicolumn{9}{||c||}{$\mu_{\rm r}^2 = \ \mu_{\rm f}^2$} \\
\multicolumn{9}{||c||}{} \\
\hline \hline
\end{tabular}
\end{center}
\end{table}

In \cref{tab:n3lo_ffns_fhmruvv,tab:n3lo_ffns_msht_fhmruvv,tab:n3lo_ffns_mshtprior,tab:n3lo_ffns_mshtposterior,tab:n3lo_ffns_nnpdf} we report the central values for the FFNS 
aN$^3$LO evolution for the 3 different splitting function approximations from FHMRUVV, MSHT and NNPDF respectively.
The first part of the table refers to the central scale $\mu_{\rm f} = \mu_{\rm r}$ while the second and the third parts refer to the scale varied results. For MSHT we report only the values at the central scale $\mu_{\rm f} = \mu_{\rm r}$.
Taking the same splitting function approximations from FHMRUVV as input, \cref{tab:n3lo_ffns_msht_fhmruvv,tab:n3lo_ffns_fhmruvv} 
show a comparision of the two aN$^3$LO evolution codes used by MSHT20 and NNPDF in the FFNS scheme at the central scale $\mu_{\rm f} = \mu_{\rm r}$.
In the range $10^{-5} \lesssim x \lesssim 0.5$ there is generally very good agreement between the two evolution codes, with slightly larger deviations observed in regions of $x$ where PDFs are becoming very small, i.e., 
in the limits $x \to 0$ (vanishing valence PDFs) and $x \to 1$.
This very good agreement among the evolution codes provides the basis for quantifying the aN$^3$LO FFNS evolution, now with the 
splitting function approximations from MSHT20 as input (\cref{tab:n3lo_ffns_mshtprior} with MSHT20 prior and \cref{tab:n3lo_ffns_mshtposterior} with MSHT20 posterior) using the MSHT20 code, and for the NNPDF approximations (\cref{tab:n3lo_ffns_nnpdf}) with the NNPDF code, cf.~\cref{input+tools} for details.
Due to the different methodology in perparing those approximations, larger deviations are observed as a result of the aN$^3$LO evolution, especially for 
smaller $x$ values, e.g., for the gluon PDF at $x \lesssim 10^{-3}$, when sufficient constraints on the splitting fuction $P_{gg}^{\,(3)}(x)$ from Mellin moments are lacking. 
In essence, these deviations demonstrate the current uncertainties inherent in the knowledge of the evolution kernels. However, these deviations would reduce with the use of more up to date input information for each of the groups determination of the splitting functions (particularly for the older MSHT determination). 
%{\color{blue}
%Added a line here - RST}
%

The corresponding values for the VFNS evolution are listed in
\cref{tab:n3lo_vfns_fhmruvv,tab:n3lo_vfns_fhmruvv_msht,tab:n3lo_vfns_msht_prior,tab:n3lo_vfns_msht_posterior,tab:n3lo_vfns_nnpdf}.
\cref{tab:n3lo_vfns_fhmruvv,tab:n3lo_vfns_fhmruvv_msht} show again very good agreement 
in the same $x$ range as before between the two aN$^3$LO evolution codes (MSHT20 and NNPDF) with the FHMRUVV splitting function approximations, now employing the VFNS scheme, while 
\cref{tab:n3lo_vfns_msht_prior,tab:n3lo_vfns_msht_posterior} 
(MSHT20 prior and posterior with the MSHT20 code) and 
\cref{tab:n3lo_ffns_nnpdf} (NNPDF approximations with NNPDF code) in the VFNS scheme display the same similarities and differences as observed in the FFNS scheme.

For simplicity we only give the central values in \cref{tab:n3lo_ffns_fhmruvv,tab:n3lo_ffns_msht_fhmruvv,tab:n3lo_ffns_mshtprior,tab:n3lo_ffns_mshtposterior,tab:n3lo_ffns_nnpdf,tab:n3lo_vfns_fhmruvv,tab:n3lo_vfns_fhmruvv_msht,tab:n3lo_vfns_msht_prior,tab:n3lo_vfns_msht_posterior,tab:n3lo_vfns_nnpdf}, 
while the complete set containing all evolved PDFs with all the splitting functions variations are available under the following URL:
\begin{center}
\url{https://www.hep.ucl.ac.uk/pdf4lhc/aN3LObenchmarking.shtml}
\end{center}

\begin{figure}
    \centering
    \includegraphics[width=\textwidth, height=0.95\textheight]
    {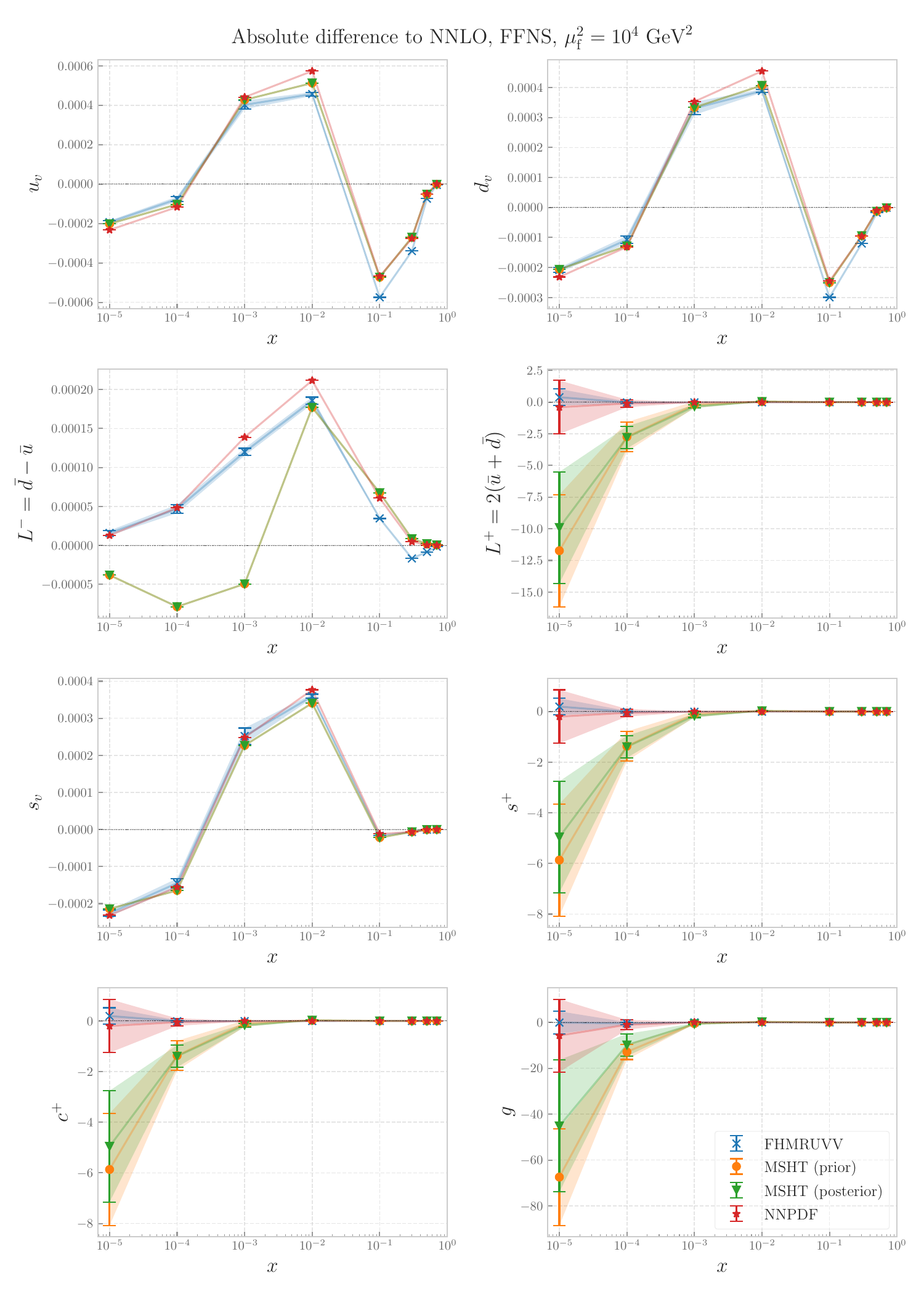}
    \caption{
        Absolute difference of the aN$^3$LO evolution with respect to the NNLO, for the initial conditions and the input parton distributions
        given in Sec.~\ref{sec:toy_pdf} with the FFNS settings.
        We display results for FHMRUVV (blue), MSHT prior (orange),  MSHT posterior (green) and NNPDF (red) approximations.
        The displayed FHMRUVV result is averaged between the values of the MSHT20 and NNPDF evolution code.
    }
    \label{fig:n3lo_bench_ffns_abs}
\end{figure}

\begin{figure}
    \centering
    \includegraphics[width=\textwidth, height=0.95\textheight]{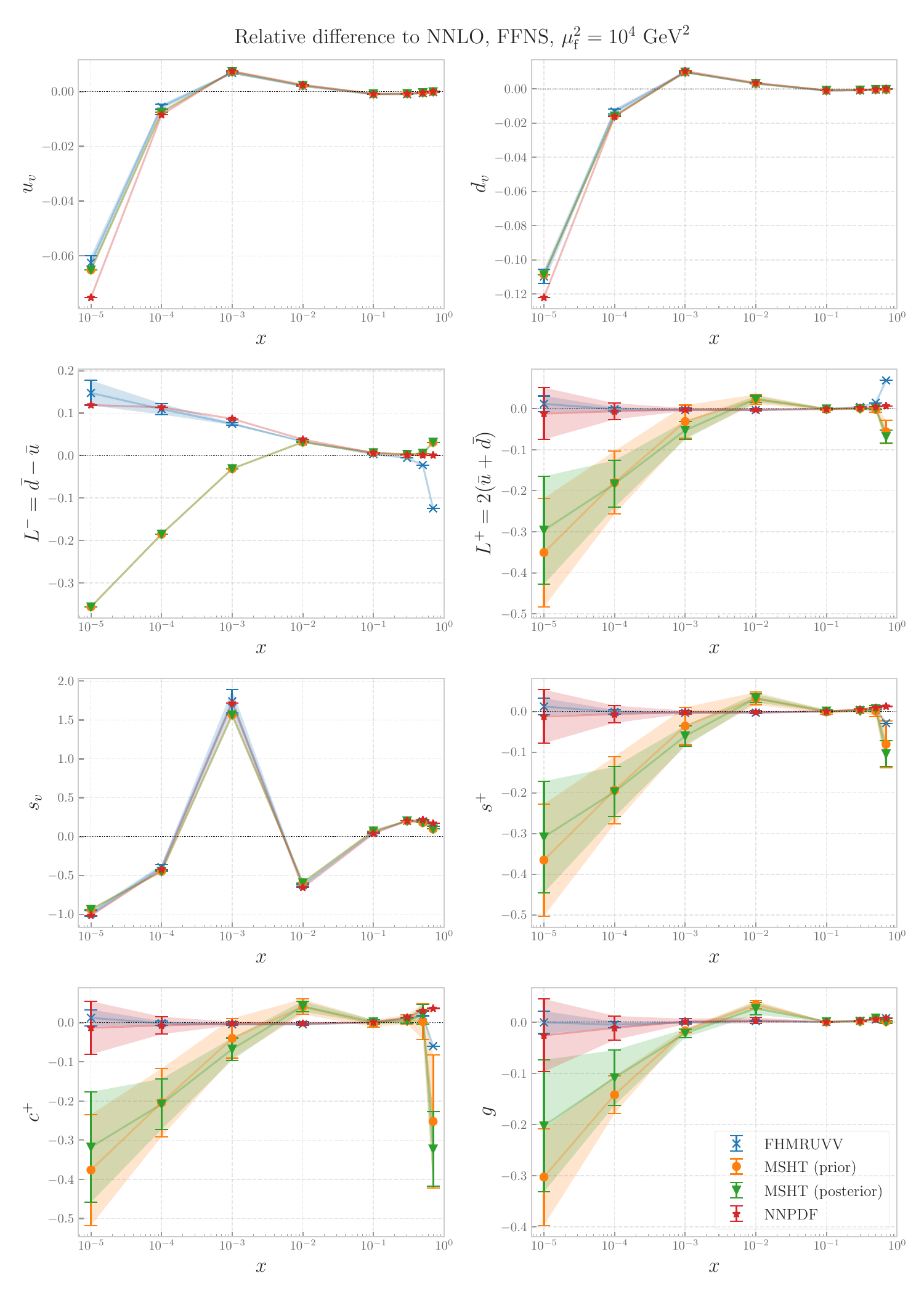}
    \caption{Same as \cref{fig:n3lo_bench_ffns_abs}, but now displaying the relative difference with respect to the NNLO evolution.}
    \label{fig:n3lo_bench_ffns_rel}
\end{figure}

\begin{figure}
    \centering
    \includegraphics[width=\textwidth, height=0.95\textheight]
    {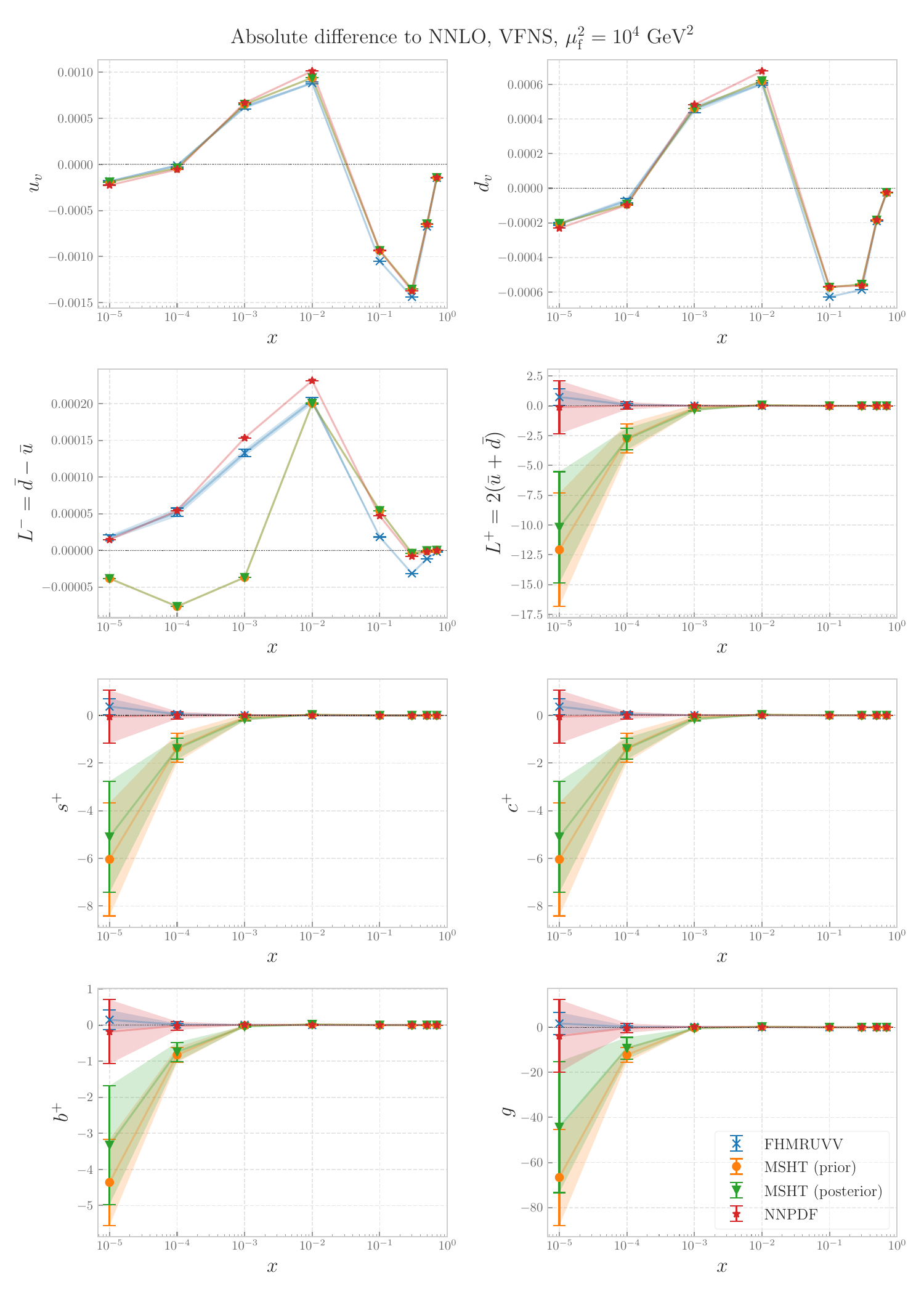}
    \caption{Same as \cref{fig:n3lo_bench_ffns_abs}, but now for VFNS evolution settings.}
    \label{fig:n3lo_bench_vfns_abs}
\end{figure}

\begin{figure}
    \centering
    \includegraphics[width=\textwidth, height=0.95\textheight]{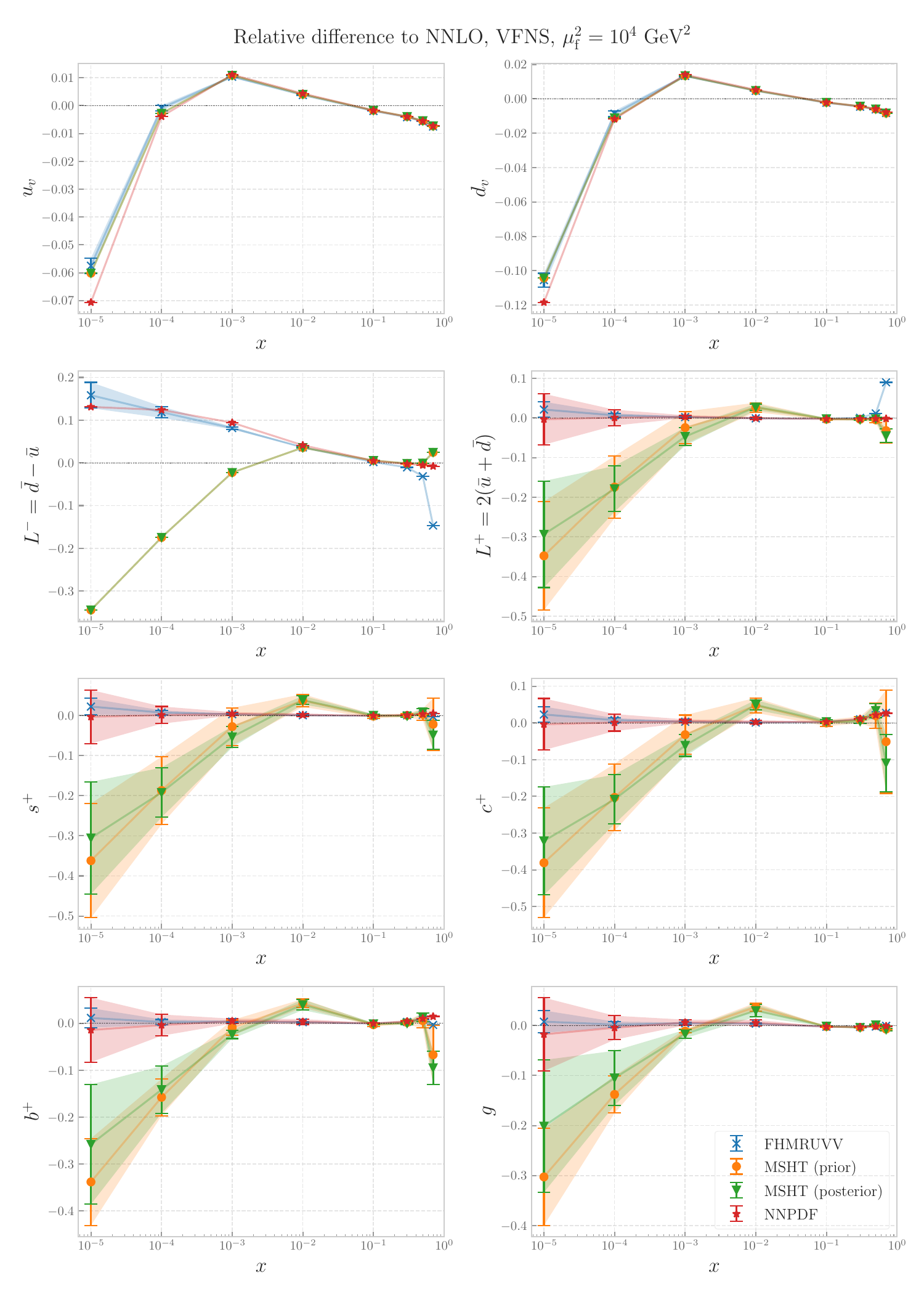}
    \caption{Same as \cref{fig:n3lo_bench_ffns_rel}, but now for VFNS evolution settings.}
    \label{fig:n3lo_bench_vfns_rel}
\end{figure}

The benchmark numbers are illustrated in \cref{fig:n3lo_bench_ffns_abs,fig:n3lo_bench_ffns_rel,fig:n3lo_bench_vfns_abs,fig:n3lo_bench_vfns_rel}, where we show the absolute and relative difference with respect to the NNLO evolution for the different aN$^3$LO evolutions, for each PDF flavor combination, reported in the previous tables. We display the FFNS and VFNS settings respectively with the central scale $\mu_{\rm r} = \mu_{\rm f}$.
The figures include an estimate of the uncertainties due to the approximated splitting functions, which is obtained by varying a single splitting function at a time during the evolution, and taking the uncertainty, for each point in $x$, as $1/2$ of the spread of the final PDF.
The displayed FHMRUVV result is averaged between the values obtained with the MSHT20 and NNPDF evolution codes. 
Since MSHT and NNPDF parametrizations do not include uncertainties for the non-singlet combinations, errors for the combinations $d_v, u_v$, $s_V$ and $L_{-}$ have been set to 0, though the difference from different approximations (which is largely at small $x$) can be regarded as a measure of the uncertainty. This is generally very small on an absolute scale due to the smallness of the corresponding non-singlet flavor combinations. 
\cref{fig:n3lo_bench_ffns_abs,fig:n3lo_bench_ffns_rel} for the evolution in the FFNS scheme show that all the N$^3$LO approximations provide consistent results in the data region of $x \gtrsim 10^{-3}$, with at most few percent differences relative to NNLO and often less for most PDF flavor combinations, whilst absolute differences are smaller still.
There are some remaining differences in the small-$x$ region where the uncertainties of the individual approximations also grow due to unknown small-$x$ logarithms and so the theoretical uncertainties are larger.
Thus in the kinematic range of the LHC, the benchmark numbers for the aN$^3$LO evolution demonstrate good perturbative convergence and a significant reduction of the residual theoretical uncertainty.
Hence, they should be sufficient for most collider-physics applications.
Similar findings are observed in \cref{fig:n3lo_bench_vfns_abs,fig:n3lo_bench_vfns_rel} for the evolution in the VFNS scheme.

\begin{figure}
    \centering
    \includegraphics[width=0.48\textwidth]{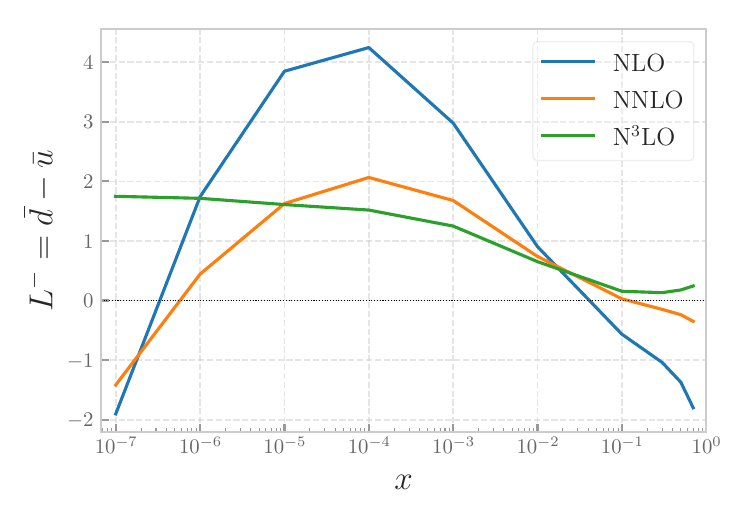}
    \includegraphics[width=0.48\textwidth]{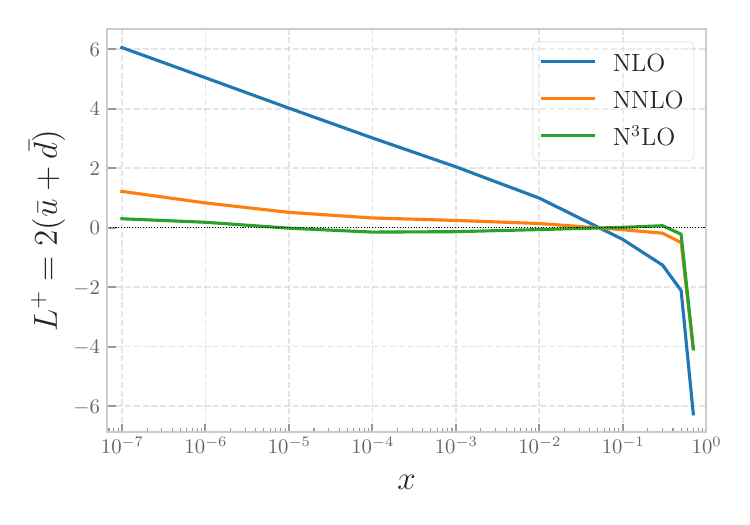} \\
    \includegraphics[width=0.48\textwidth]{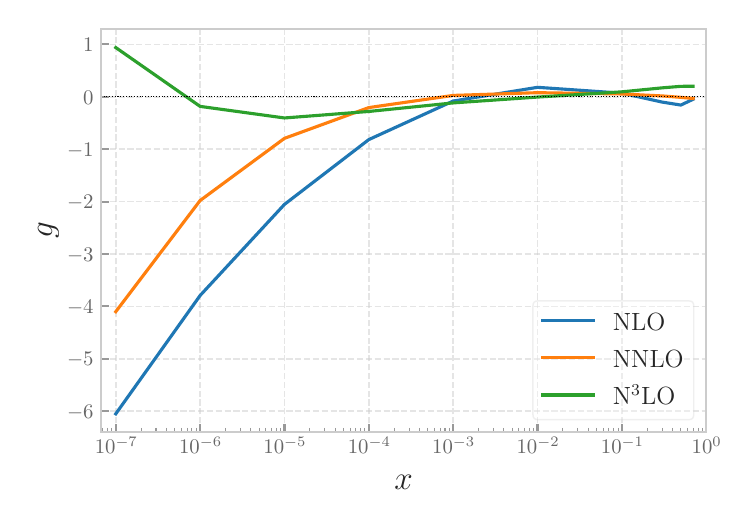}
    \caption{Relative difference in percent between the exact and truncated solutiong of the DGLAP evolution at different pertubative orders. We adopt the same evolution and boundary condition as in the benchmark exercise and display the result for the non-singlet quark combinations $L^-$, $L^+$ and the gluon $g$.}
    \label{fig:diff_trn_exa}
\end{figure}

As a final study, in order to check the stability of the DGLAP kernels
at different QCD orders, we investigate the effect of different DGLAP solution methods \cite{Vogt:2004ns}, called \textit{exact}
and \textit{truncated}, which differ by the inclusion of higher order terms. The former is utilised by MSHT in their $x$-space implementation and the latter by NNPDF in their Mellin space implementation. 
In the NNPDF code either method of solution is implemented and we utilise this to compare the impact of this difference.
From \cref{fig:diff_trn_exa} we observe that, as higher orders in the splitting functions are included the difference between the different solution methods is smaller, indicating a good perturbative convergence. This is visible both for non-singlet like distributions ($L^-$) and singlet-like ($L^+$ and $g$).

\section{Conclusions and Outlook}
\label{sec:outlook}

The study has shown excellent agreement between MSHT and NNPDF evolution when exactly the same versions of the N$^3$LO splitting functions are used.  In addition, it is observed that
the differences for $x>10^{-3}$ are very small when the distinct original estimates used in \cite{McGowan:2022nag} and \cite{NNPDF:2024nan} are used.
Moreover, the origin of the small differences in this region, and larger ones for smaller $x$  is well understood in terms of differences in the approximate splitting functions.
Hence, the study has been a success in this regard.
Furthermore, we can see in \cref{fig:diff_trn_exa} that there is also good convergence in the difference between the exact DGLAP evolution and the truncated solution method as the order of the equation used increases, and specifically that at N$^3$LO this is also small fractions of a percent except at very small $x$ or in very small non-singlet quark combinations.
Thus, overall, we can have high confidence in our precision in PDF evolution at N$^3$LO.
However, there are obvious further avenues for exploration. 

As mentioned in \cref{sec:hq}, the ${\cal O}(\alpha_S^3)$ versions of the OMEs for transition across heavy flavour transition points are now complete \cite{Ablinger:2014nga,Ablinger:2024xtt,Kawamura:2012cr,Ablinger:2022wbb,Bierenbaum:2009mv,Ablinger:2014vwa,Blumlein:2021enk,Ablinger:2014uka,Ablinger:2014tla,ablinger:agq}, though the final parts only became available while our study was already well underway. Also, not all parts are currently in an easily usable form. Hence we only included the transition matrix elements up to ${\cal O}(\alpha_S^2)$, i.e.\ NNLO in this study. In the future we can consider the effects of improving upon this. 

There are also effects of additional information~\cite{Falcioni:2023tzp,Falcioni:2023luc,Falcioni:2023vqq,Moch:2023tdj}, such as moments, computed since the original MSHT estimates of the splitting functions.
We have studied the effect of the benchmark evolution of MSHT including instead the FHMRUVV determination of the splitting functions, but it would be of at least as much interest to examine the effect of this update in a full PDF fit.
Indeed, MSHT have preliminary results \cite{Thorne:2024_DISproceedings} which show changes mainly in the gluon distribution, but remaining within the uncertainty band of the original MSHT determination. These bring the change in the MSHT aN$^3$LO gluon compared to that at NNLO somewhat closer to that observed by NNPDF, which also observes a dip in the gluon around $x \sim 0.01$ but of reduced magnitude.
To be explicit the change in the dip at $x \sim 0.01$ is reduced by about 1.5$\%$.
Finally very recently additional moments for $P_{gq}$ were determined in \cite{Falcioni:2024xyt}, these are therefore included in neither of the PDF groups approximations and so it would be interesting to see the effect on the PDFs, and corresponding improvements in agreement.
As such, the level of variation in N$^3$LO evolution illustrated in \cref{fig:n3lo_bench_ffns_abs,fig:n3lo_bench_ffns_rel,fig:n3lo_bench_vfns_abs,fig:n3lo_bench_vfns_rel} is very much an upper estimate, which will be reduced when all up to date information which can be used to estimate N$^3$LO splitting functions now available is used by all groups.
Uncertainties from other sources, e.g.\ limitations in the precision of  cross sections to use together with N$^3$LO evolution in PDF extractions, is a possible study for the future. In addition, there are also methodological differences in the implementation of the associated theoretical uncertainties.

However, given that the results in this report already show a good level of convergence in the most important regions for N$^3$LO PDF evolution phenomenological studies, we can conclude that at least as far as an understanding of this evolution is concerned, we are reaching a point where relevant uncertainties are very small and are estimated reliably. 

\section*{Acknowledgements}

T.~C. acknowledges that this project has received funding from the European Research Council (ERC) under the European Union’s Horizon 2020 research and innovation programme (Grant agreement No. 101002090 COLORFREE).  L.~H.-L. and R.S.~T. thank STFC for support via grant awards ST/T000856/1 and ST/X000516/1.
F.~H. is supported by the Academy of Finland project 358090 and is funded as a part of the Center of Excellence in Quark Matter of the Academy of Finland, project 346326.
S.~M. acknowledges the ERC Advanced Grant 101095857 Conformal-EIC.

\bibliography{a3nlobib}

\providecommand{\href}[2]{#2}\begingroup\raggedright\begin{thebibliography}{10}

\bibitem{Giele:2002hx}
W.~Giele et~al., \emph{{The QCD / SM working group: Summary report}},  \href{https://arxiv.org/abs/hep-ph/0204316}{{\ttfamily hep-ph/0204316}}.

\bibitem{Dittmar:2005ed}
M.~Dittmar et~al., \emph{{Working Group I: Parton distributions: Summary report for the HERA LHC Workshop Proceedings}},  \href{https://arxiv.org/abs/hep-ph/0511119}{{\ttfamily hep-ph/0511119}}.

\bibitem{Andersen:2024czj}
J.~Andersen et~al., \emph{{Les Houches 2023: Physics at TeV Colliders: Standard Model Working Group Report}},  in \emph{{Physics of the TeV Scale and Beyond the Standard Model}: {Intensifying the Quest for New Physics}}, 6, 2024, \href{https://arxiv.org/abs/2406.00708}{{\ttfamily 2406.00708}}.

\bibitem{Baikov:2016tgj}
P.~A. Baikov, K.~G. Chetyrkin and J.~H. K\"uhn, \emph{{Five-Loop Running of the QCD coupling constant}}, \href{https://doi.org/10.1103/PhysRevLett.118.082002}{\emph{Phys. Rev. Lett.} {\bfseries 118} (2017) 082002} [\href{https://arxiv.org/abs/1606.08659}{{\ttfamily 1606.08659}}].

\bibitem{Herzog:2017ohr}
F.~Herzog, B.~Ruijl, T.~Ueda, J.~A.~M. Vermaseren and A.~Vogt, \emph{{The five-loop beta function of Yang-Mills theory with fermions}}, \href{https://doi.org/10.1007/JHEP02(2017)090}{\emph{JHEP} {\bfseries 02} (2017) 090} [\href{https://arxiv.org/abs/1701.01404}{{\ttfamily 1701.01404}}].

\bibitem{Chetyrkin:2017bjc}
K.~G. Chetyrkin, G.~Falcioni, F.~Herzog and J.~A.~M. Vermaseren, \emph{{Five-loop renormalisation of QCD in covariant gauges}}, \href{https://doi.org/10.1007/JHEP10(2017)179}{\emph{JHEP} {\bfseries 10} (2017) 179} [\href{https://arxiv.org/abs/1709.08541}{{\ttfamily 1709.08541}}].

\bibitem{Luthe:2017ttg}
T.~Luthe, A.~Maier, P.~Marquard and Y.~Schroder, \emph{{The five-loop Beta function for a general gauge group and anomalous dimensions beyond Feynman gauge}}, \href{https://doi.org/10.1007/JHEP10(2017)166}{\emph{JHEP} {\bfseries 10} (2017) 166} [\href{https://arxiv.org/abs/1709.07718}{{\ttfamily 1709.07718}}].

\bibitem{Moch:2004pa}
S.~Moch, J.~A.~M. Vermaseren and A.~Vogt, \emph{{The Three loop splitting functions in QCD: The Nonsinglet case}}, \href{https://doi.org/10.1016/j.nuclphysb.2004.03.030}{\emph{Nucl. Phys. B} {\bfseries 688} (2004) 101} [\href{https://arxiv.org/abs/hep-ph/0403192}{{\ttfamily hep-ph/0403192}}].

\bibitem{Vogt:2004mw}
A.~Vogt, S.~Moch and J.~A.~M. Vermaseren, \emph{{The Three-loop splitting functions in QCD: The Singlet case}}, \href{https://doi.org/10.1016/j.nuclphysb.2004.04.024}{\emph{Nucl. Phys. B} {\bfseries 691} (2004) 129} [\href{https://arxiv.org/abs/hep-ph/0404111}{{\ttfamily hep-ph/0404111}}].

\bibitem{Moch:2017uml}
S.~Moch, B.~Ruijl, T.~Ueda, J.~A.~M. Vermaseren and A.~Vogt, \emph{{Four-Loop Non-Singlet Splitting Functions in the Planar Limit and Beyond}}, \href{https://doi.org/10.1007/JHEP10(2017)041}{\emph{JHEP} {\bfseries 10} (2017) 041} [\href{https://arxiv.org/abs/1707.08315}{{\ttfamily 1707.08315}}].

\bibitem{Davies:2016jie}
J.~Davies, A.~Vogt, B.~Ruijl, T.~Ueda and J.~A.~M. Vermaseren, \emph{{Large-nf contributions to the four-loop splitting functions in QCD}}, \href{https://doi.org/10.1016/j.nuclphysb.2016.12.012}{\emph{Nucl. Phys. B} {\bfseries 915} (2017) 335} [\href{https://arxiv.org/abs/1610.07477}{{\ttfamily 1610.07477}}].

\bibitem{Gehrmann:2023cqm}
T.~Gehrmann, A.~von Manteuffel, V.~Sotnikov and T.-Z. Yang, \emph{{Complete $ {N}_f^2 $ contributions to four-loop pure-singlet splitting functions}}, \href{https://doi.org/10.1007/JHEP01(2024)029}{\emph{JHEP} {\bfseries 01} (2024) 029} [\href{https://arxiv.org/abs/2308.07958}{{\ttfamily 2308.07958}}].

\bibitem{Gehrmann:2023iah}
T.~Gehrmann, A.~von Manteuffel, V.~Sotnikov and T.-Z. Yang, \emph{{The $N_fC_F^3$ contribution to the non-singlet splitting function at four-loop order}}, \href{https://doi.org/10.1016/j.physletb.2023.138427}{\emph{Phys. Lett. B} {\bfseries 849} (2024) 138427} [\href{https://arxiv.org/abs/2310.12240}{{\ttfamily 2310.12240}}].

\bibitem{Falcioni:2023tzp}
G.~Falcioni, F.~Herzog, S.~Moch, J.~Vermaseren and A.~Vogt, \emph{{The double fermionic contribution to the four-loop quark-to-gluon splitting function}}, \href{https://doi.org/10.1016/j.physletb.2023.138351}{\emph{Phys. Lett. B} {\bfseries 848} (2024) 138351} [\href{https://arxiv.org/abs/2310.01245}{{\ttfamily 2310.01245}}].

\bibitem{Moch:2018wjh}
S.~Moch, B.~Ruijl, T.~Ueda, J.~A.~M. Vermaseren and A.~Vogt, \emph{{On quartic colour factors in splitting functions and the gluon cusp anomalous dimension}}, \href{https://doi.org/10.1016/j.physletb.2018.06.017}{\emph{Phys. Lett. B} {\bfseries 782} (2018) 627} [\href{https://arxiv.org/abs/1805.09638}{{\ttfamily 1805.09638}}].

\bibitem{Moch:2021qrk}
S.~Moch, B.~Ruijl, T.~Ueda, J.~A.~M. Vermaseren and A.~Vogt, \emph{{Low moments of the four-loop splitting functions in QCD}}, \href{https://doi.org/10.1016/j.physletb.2021.136853}{\emph{Phys. Lett. B} {\bfseries 825} (2022) 136853} [\href{https://arxiv.org/abs/2111.15561}{{\ttfamily 2111.15561}}].

\bibitem{Falcioni:2023luc}
G.~Falcioni, F.~Herzog, S.~Moch and A.~Vogt, \emph{{Four-loop splitting functions in QCD \textendash{} The quark-quark case}}, \href{https://doi.org/10.1016/j.physletb.2023.137944}{\emph{Phys. Lett. B} {\bfseries 842} (2023) 137944} [\href{https://arxiv.org/abs/2302.07593}{{\ttfamily 2302.07593}}].

\bibitem{Falcioni:2023vqq}
G.~Falcioni, F.~Herzog, S.~Moch and A.~Vogt, \emph{{Four-loop splitting functions in QCD \textendash{} The gluon-to-quark case}}, \href{https://doi.org/10.1016/j.physletb.2023.138215}{\emph{Phys. Lett. B} {\bfseries 846} (2023) 138215} [\href{https://arxiv.org/abs/2307.04158}{{\ttfamily 2307.04158}}].

\bibitem{Moch:2023tdj}
S.~Moch, B.~Ruijl, T.~Ueda, J.~Vermaseren and A.~Vogt, \emph{{Additional moments and x-space approximations of four-loop splitting functions in QCD}},  \href{https://arxiv.org/abs/2310.05744}{{\ttfamily 2310.05744}}.

\bibitem{Falcioni:2024xyt}
G.~Falcioni, F.~Herzog, S.~Moch, A.~Pelloni and A.~Vogt, \emph{{Four-loop splitting functions in QCD -- The quark-to-gluon case}},  \href{https://arxiv.org/abs/2404.09701}{{\ttfamily 2404.09701}}.

\bibitem{Schroder:2005hy}
Y.~Schr{\"o}der and M.~Steinhauser, \emph{{Four-loop decoupling relations for the strong coupling}}, \href{https://doi.org/10.1088/1126-6708/2006/01/051}{\emph{JHEP} {\bfseries 01} (2006) 051} [\href{https://arxiv.org/abs/hep-ph/0512058}{{\ttfamily hep-ph/0512058}}].

\bibitem{Chetyrkin:2005ia}
K.~G. Chetyrkin, J.~H. K{\"u}hn and C.~Sturm, \emph{{QCD decoupling at four loops}}, \href{https://doi.org/10.1016/j.nuclphysb.2006.03.020}{\emph{Nucl. Phys. B} {\bfseries 744} (2006) 121} [\href{https://arxiv.org/abs/hep-ph/0512060}{{\ttfamily hep-ph/0512060}}].

\bibitem{Buza:1995ie}
M.~Buza, Y.~Matiounine, J.~Smith, R.~Migneron and W.~L. van Neerven, \emph{{Heavy quark coefficient functions at asymptotic values $Q^2 \gg m^2$}}, \href{https://doi.org/10.1016/0550-3213(96)00228-3}{\emph{Nucl. Phys. B} {\bfseries 472} (1996) 611} [\href{https://arxiv.org/abs/hep-ph/9601302}{{\ttfamily hep-ph/9601302}}].

\bibitem{Buza:1998wv}
M.~Buza, Y.~Matiounine, J.~Smith and W.~L. van Neerven, \emph{{Charm electroproduction viewed in the variable flavor number scheme versus fixed order perturbation theory}}, \href{https://doi.org/10.1007/BF01245820}{\emph{Eur. Phys. J. C} {\bfseries 1} (1998) 301} [\href{https://arxiv.org/abs/hep-ph/9612398}{{\ttfamily hep-ph/9612398}}].

\bibitem{Ablinger:2010ty}
J.~Ablinger, J.~Blumlein, S.~Klein, C.~Schneider and F.~Wissbrock, \emph{{The $O(\alpha_s^3)$ Massive Operator Matrix Elements of $O(n_f)$ for the Structure Function $F_2(x,Q^2)$ and Transversity}}, \href{https://doi.org/10.1016/j.nuclphysb.2010.10.021}{\emph{Nucl. Phys. B} {\bfseries 844} (2011) 26} [\href{https://arxiv.org/abs/1008.3347}{{\ttfamily 1008.3347}}].

\bibitem{Ablinger:2014lka}
J.~Ablinger, J.~Blümlein, A.~De~Freitas, A.~Hasselhuhn, A.~von Manteuffel, M.~Round et~al., \emph{The transition matrix element $a_{gq}(n)$ of the variable flavor number scheme at $\cal{O}(\alpha_s^3)$}, \href{https://doi.org/10.1016/j.nuclphysb.2014.02.007}{\emph{Nuclear Physics B} {\bfseries 882} (2014) 263–288}.

\bibitem{Ablinger:2014nga}
J.~Ablinger, A.~Behring, J.~Bl\"umlein, A.~De~Freitas, A.~von Manteuffel and C.~Schneider, \emph{{The 3-loop pure singlet heavy flavor contributions to the structure function $F_2(x,Q^2)$ and the anomalous dimension}}, \href{https://doi.org/10.1016/j.nuclphysb.2014.10.008}{\emph{Nucl. Phys. B} {\bfseries 890} (2014) 48} [\href{https://arxiv.org/abs/1409.1135}{{\ttfamily 1409.1135}}].

\bibitem{Ablinger:2023ahe}
J.~Ablinger, A.~Behring, J.~Bl\"umlein, A.~De~Freitas, A.~von Manteuffel, C.~Schneider et~al., \emph{{The first\textendash{}order factorizable contributions to the three\textendash{}loop massive operator matrix elements $A_{Qg}^{(3)}$ and $\ensuremath{\Delta}A_{Qg}^{(3)}$}}, \href{https://doi.org/10.1016/j.nuclphysb.2023.116427}{\emph{Nucl. Phys. B} {\bfseries 999} (2024) 116427} [\href{https://arxiv.org/abs/2311.00644}{{\ttfamily 2311.00644}}].

\bibitem{Ablinger:2024xtt}
J.~Ablinger, A.~Behring, J.~Bl\"umlein, A.~De~Freitas, A.~von Manteuffel, C.~Schneider et~al., \emph{{The non-first-order-factorizable contributions to the three-loop single-mass operator matrix elements $A_{Qg}^{(3)}$ and $\Delta A_{Qg}^{(3)}$}},  \href{https://arxiv.org/abs/2403.00513}{{\ttfamily 2403.00513}}.

\bibitem{Kawamura:2012cr}
H.~Kawamura, N.~A. Lo~Presti, S.~Moch and A.~Vogt, \emph{{On the next-to-next-to-leading order QCD corrections to heavy-quark production in deep-inelastic scattering}}, \href{https://doi.org/10.1016/j.nuclphysb.2012.07.001}{\emph{Nucl. Phys. B} {\bfseries 864} (2012) 399} [\href{https://arxiv.org/abs/1205.5727}{{\ttfamily 1205.5727}}].

\bibitem{Alekhin:2017kpj}
S.~Alekhin, J.~Bl\"umlein, S.~Moch and R.~Placakyte, \emph{{Parton distribution functions, $\alpha_s$, and heavy-quark masses for LHC Run II}}, \href{https://doi.org/10.1103/PhysRevD.96.014011}{\emph{Phys. Rev. D} {\bfseries 96} (2017) 014011} [\href{https://arxiv.org/abs/1701.05838}{{\ttfamily 1701.05838}}].

\bibitem{Buza:1996wv}
M.~Buza, Y.~Matiounine, J.~Smith and W.~L. van Neerven, \emph{{Charm electroproduction viewed in the variable flavor number scheme versus fixed order perturbation theory}}, \href{https://doi.org/10.1007/BF01245820}{\emph{Eur. Phys. J. C} {\bfseries 1} (1998) 301} [\href{https://arxiv.org/abs/hep-ph/9612398}{{\ttfamily hep-ph/9612398}}].

\bibitem{Ablinger:2017xml}
J.~Ablinger, J.~Bl\"umlein, A.~De~Freitas, C.~Schneider and K.~Sch\"onwald, \emph{{The two-mass contribution to the three-loop pure singlet operator matrix element}}, \href{https://doi.org/10.1016/j.nuclphysb.2017.12.018}{\emph{Nucl. Phys. B} {\bfseries 927} (2018) 339} [\href{https://arxiv.org/abs/1711.06717}{{\ttfamily 1711.06717}}].

\bibitem{Ablinger:2018brx}
J.~Ablinger, J.~Bl\"umlein, A.~De~Freitas, A.~Goedicke, C.~Schneider and K.~Sch\"onwald, \emph{{The Two-mass Contribution to the Three-Loop Gluonic Operator Matrix Element $A_{gg,Q}^{(3)}$}}, \href{https://doi.org/10.1016/j.nuclphysb.2018.04.023}{\emph{Nucl. Phys. B} {\bfseries 932} (2018) 129} [\href{https://arxiv.org/abs/1804.02226}{{\ttfamily 1804.02226}}].

\bibitem{Ablinger:2022wbb}
J.~Ablinger, A.~Behring, J.~Bl\"umlein, A.~De~Freitas, A.~Goedicke, A.~von Manteuffel et~al., \emph{{The unpolarized and polarized single-mass three-loop heavy flavor operator matrix elements A$_{gg,Q}$ and \ensuremath{\Delta}A$_{gg,Q}$}}, \href{https://doi.org/10.1007/JHEP12(2022)134}{\emph{JHEP} {\bfseries 12} (2022) 134} [\href{https://arxiv.org/abs/2211.05462}{{\ttfamily 2211.05462}}].

\bibitem{Ablinger:2017err}
J.~Ablinger, J.~Bl\"umlein, A.~De~Freitas, A.~Hasselhuhn, C.~Schneider and F.~Wi\ss{}brock, \emph{{Three Loop Massive Operator Matrix Elements and Asymptotic Wilson Coefficients with Two Different Masses}}, \href{https://doi.org/10.1016/j.nuclphysb.2017.05.017}{\emph{Nucl. Phys. B} {\bfseries 921} (2017) 585} [\href{https://arxiv.org/abs/1705.07030}{{\ttfamily 1705.07030}}].

\bibitem{Bertone:2013vaa}
{\scshape APFEL} collaboration, \emph{{APFEL: A PDF Evolution Library with QED corrections}}, \href{https://doi.org/10.1016/j.cpc.2014.03.007}{\emph{Comput. Phys. Commun.} {\bfseries 185} (2014) 1647} [\href{https://arxiv.org/abs/1310.1394}{{\ttfamily 1310.1394}}].

\bibitem{Bertone:2017gds}
V.~Bertone, \emph{{APFEL++: A new PDF evolution library in C++}}, \href{https://doi.org/10.22323/1.297.0201}{\emph{PoS} {\bfseries DIS2017} (2018) 201} [\href{https://arxiv.org/abs/1708.00911}{{\ttfamily 1708.00911}}].

\bibitem{Candido:2022tld}
A.~Candido, F.~Hekhorn and G.~Magni, \emph{{EKO: evolution kernel operators}}, \href{https://doi.org/10.1140/epjc/s10052-022-10878-w}{\emph{Eur. Phys. J. C} {\bfseries 82} (2022) 976} [\href{https://arxiv.org/abs/2202.02338}{{\ttfamily 2202.02338}}].

\bibitem{Salam:2008qg}
G.~P. Salam and J.~Rojo, \emph{{A Higher Order Perturbative Parton Evolution Toolkit (HOPPET)}}, \href{https://doi.org/10.1016/j.cpc.2008.08.010}{\emph{Comput. Phys. Commun.} {\bfseries 180} (2009) 120} [\href{https://arxiv.org/abs/0804.3755}{{\ttfamily 0804.3755}}].

\bibitem{Vogt:2004ns}
A.~Vogt, \emph{{Efficient evolution of unpolarized and polarized parton distributions with QCD-PEGASUS}}, \href{https://doi.org/10.1016/j.cpc.2005.03.103}{\emph{Comput. Phys. Commun.} {\bfseries 170} (2005) 65} [\href{https://arxiv.org/abs/hep-ph/0408244}{{\ttfamily hep-ph/0408244}}].

\bibitem{Botje:2010ay}
M.~Botje, \emph{{QCDNUM: Fast QCD Evolution and Convolution}}, \href{https://doi.org/10.1016/j.cpc.2010.10.020}{\emph{Comput. Phys. Commun.} {\bfseries 182} (2011) 490} [\href{https://arxiv.org/abs/1005.1481}{{\ttfamily 1005.1481}}].

\bibitem{McGowan:2022nag}
J.~McGowan, T.~Cridge, L.~A. Harland-Lang and R.~S. Thorne, \emph{{Approximate N$^{3}$LO parton distribution functions with theoretical uncertainties: MSHT20aN$^3$LO PDFs}}, \href{https://doi.org/10.1140/epjc/s10052-023-11236-0}{\emph{Eur. Phys. J. C} {\bfseries 83} (2023) 185} [\href{https://arxiv.org/abs/2207.04739}{{\ttfamily 2207.04739}}].

\bibitem{Cridge:2023ryv}
T.~Cridge, L.~A. Harland-Lang and R.~S. Thorne, \emph{{Combining QED and Approximate N${}^3$LO QCD Corrections in a Global PDF Fit: MSHT20qed\_an3lo PDFs}},  \href{https://arxiv.org/abs/2312.07665}{{\ttfamily 2312.07665}}.

\bibitem{Hekhorn:2023gul}
F.~Hekhorn and G.~Magni, \emph{{DGLAP evolution of parton distributions at approximate N$^3$LO}},  \href{https://arxiv.org/abs/2306.15294}{{\ttfamily 2306.15294}}.

\bibitem{NNPDF:2024nan}
{\scshape NNPDF} collaboration, \emph{{The Path to N$^3$LO Parton Distributions}},  \href{https://arxiv.org/abs/2402.18635}{{\ttfamily 2402.18635}}.

\bibitem{Dokshitzer:2005bf}
Y.~L. Dokshitzer, G.~Marchesini and G.~P. Salam, \emph{{Revisiting parton evolution and the large-x limit}}, \href{https://doi.org/10.1016/j.physletb.2006.02.023}{\emph{Phys. Lett. B} {\bfseries 634} (2006) 504} [\href{https://arxiv.org/abs/hep-ph/0511302}{{\ttfamily hep-ph/0511302}}].

\bibitem{Fadin:1975cb}
V.~S. Fadin, E.~A. Kuraev and L.~N. Lipatov, \emph{{On the Pomeranchuk Singularity in Asymptotically Free Theories}}, \href{https://doi.org/10.1016/0370-2693(75)90524-9}{\emph{Phys. Lett. B} {\bfseries 60} (1975) 50}.

\bibitem{Kuraev:1976ge}
E.~A. Kuraev, L.~N. Lipatov and V.~S. Fadin, \emph{{Multi - Reggeon Processes in the Yang-Mills Theory}}, {\emph{Sov. Phys. JETP} {\bfseries 44} (1976) 443}.

\bibitem{Lipatov:1976zz}
L.~N. Lipatov, \emph{{Reggeization of the Vector Meson and the Vacuum Singularity in Nonabelian Gauge Theories}}, {\emph{Sov. J. Nucl. Phys.} {\bfseries 23} (1976) 338}.

\bibitem{Kuraev:1977fs}
E.~A. Kuraev, L.~N. Lipatov and V.~S. Fadin, \emph{{The Pomeranchuk Singularity in Nonabelian Gauge Theories}}, {\emph{Sov. Phys. JETP} {\bfseries 45} (1977) 199}.

\bibitem{Fadin:1998py}
V.~S. Fadin and L.~N. Lipatov, \emph{{BFKL pomeron in the next-to-leading approximation}}, \href{https://doi.org/10.1016/S0370-2693(98)00473-0}{\emph{Phys. Lett. B} {\bfseries 429} (1998) 127} [\href{https://arxiv.org/abs/hep-ph/9802290}{{\ttfamily hep-ph/9802290}}].

\bibitem{Jaroszewicz:1982gr}
T.~Jaroszewicz, \emph{{Gluonic Regge Singularities and Anomalous Dimensions in QCD}}, \href{https://doi.org/10.1016/0370-2693(82)90345-8}{\emph{Phys. Lett. B} {\bfseries 116} (1982) 291}.

\bibitem{Ciafaloni:1998gs}
M.~Ciafaloni and G.~Camici, \emph{{Energy scale(s) and next-to-leading BFKL equation}}, \href{https://doi.org/10.1016/S0370-2693(98)00551-6}{\emph{Phys. Lett. B} {\bfseries 430} (1998) 349} [\href{https://arxiv.org/abs/hep-ph/9803389}{{\ttfamily hep-ph/9803389}}].

\bibitem{Catani:1994sq}
S.~Catani and F.~Hautmann, \emph{{High-energy factorization and small x deep inelastic scattering beyond leading order}}, \href{https://doi.org/10.1016/0550-3213(94)90636-X}{\emph{Nucl. Phys. B} {\bfseries 427} (1994) 475} [\href{https://arxiv.org/abs/hep-ph/9405388}{{\ttfamily hep-ph/9405388}}].

\bibitem{Cridge:2023ozx}
T.~Cridge, L.~A. Harland-Lang and R.~S. Thorne, \emph{{The impact of LHC jet and Z$p_T$ data at up to approximate N${}^3$LO order in the MSHT global PDF fit}}, \href{https://doi.org/10.1140/epjc/s10052-024-12771-0}{\emph{Eur. Phys. J. C} {\bfseries 84} (2024) 446} [\href{https://arxiv.org/abs/2312.12505}{{\ttfamily 2312.12505}}].

\bibitem{Cridge:2024exf}
T.~Cridge, L.~A. Harland-Lang and R.~S. Thorne, \emph{{A first determination of the strong coupling $\alpha_S$ at approximate N$^{3}$LO order in a global PDF fit}},  \href{https://arxiv.org/abs/2404.02964}{{\ttfamily 2404.02964}}.

\bibitem{Barontini:2023vmr}
A.~Barontini, A.~Candido, J.~M. Cruz-Martinez, F.~Hekhorn and C.~Schwan, \emph{{Pineline: Industrialization of High-Energy Theory Predictions}},  \href{https://arxiv.org/abs/2302.12124}{{\ttfamily 2302.12124}}.

\bibitem{Henn:2019swt}
J.~M. Henn, G.~P. Korchemsky and B.~Mistlberger, \emph{{The full four-loop cusp anomalous dimension in $\mathcal{N}=4$ super Yang-Mills and QCD}}, \href{https://doi.org/10.1007/JHEP04(2020)018}{\emph{JHEP} {\bfseries 04} (2020) 018} [\href{https://arxiv.org/abs/1911.10174}{{\ttfamily 1911.10174}}].

\bibitem{Duhr:2022cob}
C.~Duhr, B.~Mistlberger and G.~Vita, \emph{{Soft integrals and soft anomalous dimensions at N$^{3}$LO and beyond}}, \href{https://doi.org/10.1007/JHEP09(2022)155}{\emph{JHEP} {\bfseries 09} (2022) 155} [\href{https://arxiv.org/abs/2205.04493}{{\ttfamily 2205.04493}}].

\bibitem{Almasy:2010wn}
A.~A. Almasy, G.~Soar and A.~Vogt, \emph{{Generalized double-logarithmic large-x resummation in inclusive deep-inelastic scattering}}, \href{https://doi.org/10.1007/JHEP03(2011)030}{\emph{JHEP} {\bfseries 03} (2011) 030} [\href{https://arxiv.org/abs/1012.3352}{{\ttfamily 1012.3352}}].

\bibitem{Davies:2022ofz}
J.~Davies, C.~H. Kom, S.~Moch and A.~Vogt, \emph{{Resummation of small-x double logarithms in QCD: inclusive deep-inelastic scattering}}, \href{https://doi.org/10.1007/JHEP08(2022)135}{\emph{JHEP} {\bfseries 08} (2022) 135} [\href{https://arxiv.org/abs/2202.10362}{{\ttfamily 2202.10362}}].

\bibitem{Ball:1999sh}
R.~D. Ball and S.~Forte, \emph{{The Small x behavior of Altarelli-Parisi splitting functions}}, \href{https://doi.org/10.1016/S0370-2693(99)01013-8}{\emph{Phys. Lett. B} {\bfseries 465} (1999) 271} [\href{https://arxiv.org/abs/hep-ph/9906222}{{\ttfamily hep-ph/9906222}}].

\bibitem{Bonvini:2018xvt}
M.~Bonvini and S.~Marzani, \emph{{Four-loop splitting functions at small $x$}}, \href{https://doi.org/10.1007/JHEP06(2018)145}{\emph{JHEP} {\bfseries 06} (2018) 145} [\href{https://arxiv.org/abs/1805.06460}{{\ttfamily 1805.06460}}].

\bibitem{NNPDF:2019ubu}
{\scshape NNPDF} collaboration, \emph{{Parton Distributions with Theory Uncertainties: General Formalism and First Phenomenological Studies}}, \href{https://doi.org/10.1140/epjc/s10052-019-7401-4}{\emph{Eur. Phys. J. C} {\bfseries 79} (2019) 931} [\href{https://arxiv.org/abs/1906.10698}{{\ttfamily 1906.10698}}].

\bibitem{NNPDF:2024dpb}
{\scshape NNPDF} collaboration, \emph{{Determination of the theory uncertainties from missing higher orders on NNLO parton distributions with percent accuracy}},  \href{https://arxiv.org/abs/2401.10319}{{\ttfamily 2401.10319}}.

\bibitem{vonManteuffel:2020vjv}
A.~von Manteuffel, E.~Panzer and R.~M. Schabinger, \emph{{Cusp and collinear anomalous dimensions in four-loop QCD from form factors}}, \href{https://doi.org/10.1103/PhysRevLett.124.162001}{\emph{Phys. Rev. Lett.} {\bfseries 124} (2020) 162001} [\href{https://arxiv.org/abs/2002.04617}{{\ttfamily 2002.04617}}].

\bibitem{Das:2019btv}
G.~Das, S.-O. Moch and A.~Vogt, \emph{{Soft corrections to inclusive deep-inelastic scattering at four loops and beyond}}, \href{https://doi.org/10.1007/JHEP03(2020)116}{\emph{JHEP} {\bfseries 03} (2020) 116} [\href{https://arxiv.org/abs/1912.12920}{{\ttfamily 1912.12920}}].

\bibitem{Das:2020adl}
G.~Das, S.~Moch and A.~Vogt, \emph{{Approximate four-loop QCD corrections to the Higgs-boson production cross section}}, \href{https://doi.org/10.1016/j.physletb.2020.135546}{\emph{Phys. Lett. B} {\bfseries 807} (2020) 135546} [\href{https://arxiv.org/abs/2004.00563}{{\ttfamily 2004.00563}}].

\bibitem{Soar:2009yh}
G.~Soar, S.~Moch, J.~A.~M. Vermaseren and A.~Vogt, \emph{{On Higgs-exchange DIS, physical evolution kernels and fourth-order splitting functions at large x}}, \href{https://doi.org/10.1016/j.nuclphysb.2010.02.003}{\emph{Nucl. Phys. B} {\bfseries 832} (2010) 152} [\href{https://arxiv.org/abs/0912.0369}{{\ttfamily 0912.0369}}].

\bibitem{Vogt:2010cv}
A.~Vogt, \emph{{Leading logarithmic large-x resummation of off-diagonal splitting functions and coefficient functions}}, \href{https://doi.org/10.1016/j.physletb.2010.06.010}{\emph{Phys. Lett. B} {\bfseries 691} (2010) 77} [\href{https://arxiv.org/abs/1005.1606}{{\ttfamily 1005.1606}}].

\bibitem{Ciafaloni:2005cg}
M.~Ciafaloni and D.~Colferai, \emph{{Dimensional regularisation and factorisation schemes in the BFKL equation at subleading level}}, \href{https://doi.org/10.1088/1126-6708/2005/09/069}{\emph{JHEP} {\bfseries 09} (2005) 069} [\href{https://arxiv.org/abs/hep-ph/0507106}{{\ttfamily hep-ph/0507106}}].

\bibitem{Ciafaloni:2006yk}
M.~Ciafaloni, D.~Colferai, G.~P. Salam and A.~M. Stasto, \emph{{Minimal subtraction vs. physical factorisation schemes in small-x QCD}}, \href{https://doi.org/10.1016/j.physletb.2006.03.014}{\emph{Phys. Lett. B} {\bfseries 635} (2006) 320} [\href{https://arxiv.org/abs/hep-ph/0601200}{{\ttfamily hep-ph/0601200}}].

\bibitem{Diehl:2021gvs}
M.~Diehl, R.~Nagar and F.~J. Tackmann, \emph{{ChiliPDF: Chebyshev interpolation for parton distributions}}, \href{https://doi.org/10.1140/epjc/s10052-022-10223-1}{\emph{Eur. Phys. J. C} {\bfseries 82} (2022) 257} [\href{https://arxiv.org/abs/2112.09703}{{\ttfamily 2112.09703}}].

\bibitem{Bierenbaum:2009mv}
I.~Bierenbaum, J.~Blumlein and S.~Klein, \emph{{Mellin Moments of the O(alpha**3(s)) Heavy Flavor Contributions to unpolarized Deep-Inelastic Scattering at Q**2 \ensuremath{>}\ensuremath{>} m**2 and Anomalous Dimensions}}, \href{https://doi.org/10.1016/j.nuclphysb.2009.06.005}{\emph{Nucl. Phys. B} {\bfseries 820} (2009) 417} [\href{https://arxiv.org/abs/0904.3563}{{\ttfamily 0904.3563}}].

\bibitem{Ablinger:2014vwa}
J.~Ablinger, A.~Behring, J.~Bl\"umlein, A.~De~Freitas, A.~Hasselhuhn, A.~von Manteuffel et~al., \emph{{The 3-Loop Non-Singlet Heavy Flavor Contributions and Anomalous Dimensions for the Structure Function $F_2(x,Q^2)$ and Transversity}}, \href{https://doi.org/10.1016/j.nuclphysb.2014.07.010}{\emph{Nucl. Phys. B} {\bfseries 886} (2014) 733} [\href{https://arxiv.org/abs/1406.4654}{{\ttfamily 1406.4654}}].

\bibitem{Blumlein:2021enk}
J.~Bl\"umlein, P.~Marquard, C.~Schneider and K.~Sch\"onwald, \emph{{The three-loop unpolarized and polarized non-singlet anomalous dimensions from off shell operator matrix elements}}, \href{https://doi.org/10.1016/j.nuclphysb.2021.115542}{\emph{Nucl. Phys. B} {\bfseries 971} (2021) 115542} [\href{https://arxiv.org/abs/2107.06267}{{\ttfamily 2107.06267}}].

\bibitem{Ablinger:2014uka}
J.~Ablinger, J.~Bl\"umlein, A.~De~Freitas, A.~Hasselhuhn, A.~von Manteuffel, M.~Round et~al., \emph{{The $O(\alpha_s^3 T_F^2)$ Contributions to the Gluonic Operator Matrix Element}}, \href{https://doi.org/10.1016/j.nuclphysb.2014.05.028}{\emph{Nucl. Phys. B} {\bfseries 885} (2014) 280} [\href{https://arxiv.org/abs/1405.4259}{{\ttfamily 1405.4259}}].

\bibitem{Ablinger:2014tla}
J.~Ablinger, J.~Bl\"umlein, A.~De~Freitas, A.~Hasselhuhn, A.~von Manteuffel, M.~Round et~al., \emph{{3-loop Massive $O(T_F^2)$ Contributions to the DIS Operator Matrix Element $A_{gg}$}}, \href{https://doi.org/10.1016/j.nuclphysbps.2015.01.009}{\emph{Nucl. Part. Phys. Proc.} {\bfseries 258-259} (2015) 37} [\href{https://arxiv.org/abs/1409.1435}{{\ttfamily 1409.1435}}].

\bibitem{ablinger:agq}
J.~{Ablinger}, J.~{Bl{\"u}mlein}, A.~{De Freitas}, A.~{Hasselhuhn}, A.~{von Manteuffel}, M.~{Round} et~al., \emph{{The transition matrix element A$_{gq}$(N) of the variable flavor number scheme at O({\ensuremath{\alpha}}s3)}}, \href{https://doi.org/10.1016/j.nuclphysb.2014.02.007}{\emph{Nuclear Physics B} {\bfseries 882} (2014) 263} [\href{https://arxiv.org/abs/1402.0359}{{\ttfamily 1402.0359}}].

\bibitem{Thorne:2024_DISproceedings}
R.~S. Thorne, T.~Cridge and L.~A. Harland-Lang, \emph{{Updates of MSHT PDFs (to appear)}},  \href{https://arxiv.org/abs/24xx.xxxxx}{{\ttfamily 24xx.xxxxx}}.

\end{thebibliography}\endgroup
\bibliographystyle{JHEP}

\end{document}